# A Review of Ontology-Driven Big Data Analytics in Healthcare: Challenges, Tools, and Applications

Ritesh Chandra*, Sonali Agarwal, Navjot Singh, and Sadhana Tiwari
Indian Institute of Information Technology Allahabad, Prayagraj, India
rsi2022001@iiita.ac.in, sonali@iiita.ac.in, navjot@iiita.ac.in, rsi2018507@iiita.ac.in

## Abstract

Exponential growth in heterogeneous healthcare data arising from electronic health records (EHRs), medical imaging, wearable sensors, and biomedical research has accelerated the adoption of data lakes and centralized architectures capable of handling the Volume, Variety, and Velocity of Big Data for advanced analytics. However, without effective governance, these repositories risk devolving into disorganized data swamps. Ontology-driven semantic data management offers a robust solution by linking metadata to healthcare knowledge graphs, thereby enhancing semantic interoperability, improving data discoverability, and enabling expressive, domain-aware access. This review adopts a systematic research strategy, formulating key research questions and conducting a structured literature search across major academic databases, with selected studies analyzed and classified into six categories of ontology-driven healthcare analytics: (i) ontology-driven integration frameworks, (ii) semantic modeling for metadata enrichment, (iii) ontology-based data access (OBDA), (iv) basic semantic data management, (v) ontology-based reasoning for decision support, and (vi) semantic annotation for unstructured data. We further examine the integration of ontology technologies with Big Data frameworks such as Hadoop, Spark, Kafka, and so on, highlighting their combined potential to deliver scalable and intelligent healthcare analytics. For each category, recent techniques, representative case studies, technical and organizational challenges, and emerging trends such as artificial intelligence, machine learning, the Internet of Things (IoT), and real-time analytics are reviewed to guide the development of sustainable, interoperable, and high-performance healthcare data ecosystems.

## 1. Introduction

The healthcare sector is undergoing a fundamental transition, owing to the increasing voluminous, and diverse health-related information. This phenomenon, known as Big Data in Healthcare, is the systematic collection, integration, and analysis of large and complex health datasets. Unlike traditional structured medical records, these datasets are in various formats, including structured, semi-structured, and unstructured data, reflecting the complex character of modern health information [1][2]. Utilizing this data is critical for improving patient outcomes, enabling early disease detection, and lowering total healthcare expenditures [3]. In clinical environments, this includes EHRs that capture patient histories and treatment details, medical imaging modalities such as MRIs, CT scans, and X-rays that primarily consist of unstructured data, and genomic datasets that reveal biological insights. Outside hospital settings, wearable devices like smartwatches and fitness trackers enable continuous, real-time health monitoring, while patient surveys and lifestyle information provide additional contextual value [4][5].

It is quite challenging to manage data of heavy volume and diversity. Conventional systems are unable to manage such complex data. If large repositories are not well managed, they may end up being unorganized collections whose usefulness is compromised by a lack of semantic organization. Healthcare organizations must adopt Big Data Analytics as a new paradigm for data-driven care, not merely a minor enhancement, in order to overcome these challenges [6]. Forecasting patient demands, detecting health hazards early, and facilitating the transition from reactive to proactive, preventative, and customized care are all made possible by advanced analytics [1]. This increases overall system efficiency, promotes evidence-based decision-making, and improves patient safety. Additionally, analytics strengthens healthcare operations at all levels by addressing socio-technical issues such as data accuracy, human error, and workflow inefficiencies [7]. The four main categories of healthcare analytics build on one another to deliver progressively greater value:

1. Summarization-Through the summarization of medical histories, examinations, and treatment outcomes, descriptive analytics offers a fundamental perspective that sheds light on a patient's present state of health [8].
2. Trends and Connections Identification- By identifying trends and connections, diagnostic analytics explains the underlying causes of medical events and supports recurrence prevention measures. By classifying patients according to their clinical and molecular traits, it also makes precision therapy possible [9].
3. Event Forecasting- Predictive analytics forecasts future events by using statistical models and past data. Predictive models, for instance, can forecast equipment use during patient surges or estimate the likelihood of COVID-19 death based on patient demographics [10].
4. Emerging clinical recommendations with prescriptive insights- The most sophisticated type is called prescriptive analytics, which combines clinical recommendations with predictive insights to suggest targeted, doable treatments for the best possible care [11].

Together, these approaches represent a maturity model that moves from understanding past events to actively shaping future outcomes. This evolution reflects a strategic shift in healthcare from retrospective analysis to proactive intervention and continuous optimization. By equipping clinicians and administrators with timely access to both real-time and historical information, big data analytics supports informed decision-making, enhances resource utilization, and ensures more effective and cost-efficient operations. It enables personalized, preventive, and high-quality care, making a measurable difference in patient outcomes.

### 1.1. Role of Ontology in Enhancing Big Data Analytics

Health data ontologies are formal representations of knowledge that define concepts, their relationships, and governing rules within the healthcare domain. It provides a common vocabulary and a robust semantic framework, essential for seamless data sharing, integration, and advanced analysis across various healthcare systems and applications. The key contributions of ontologies in enhancing Big Data Analytics include:

1. Standardized Representation: Establishing uniform and unambiguous ways of structuring health data across heterogeneous datasets [12].
2. Data Integration and Interoperability: Enabling consistent integration of data from sources such as EHRs, claims data, and public health surveillance systems [13].
3. Semantic Reasoning and Insights: Supporting advanced analytics through reasoning and inference to uncover hidden patterns and relationships [14].

4. Decision-Making Support: Providing coherent frameworks for interpreting data, thereby aiding clinical and public health policy decisions [15].
5. Reducing Redundancy and Ambiguity: Minimizing duplication and inconsistencies in medical terminology, documentation, and coding .
6. AI and Automation Enablement: Structuring data in a machine-readable format to enhance AI-driven applications, improving accuracy and efficiency [16].
7. Semantic Bridging: Linking siloed systems (like EMRs, billing platforms, payer portals, regulatory databases) to enable unambiguous communication and smooth data flow [17].

Ontologies serve as more than static data dictionaries. They provide a semantic backbone that allows machines to interpret healthcare data meaningfully. This extends beyond syntactic compatibility to achieve true semantic interoperability, which is an essential capability for executing complex analytics and integrating artificial intelligence effectively. By explicitly structuring relationships among healthcare concepts, ontologies optimize data flow, reduce errors, and support high-quality clinical decision-making. The absence of robust ontological frameworks increase persistent issues in healthcare, including data silos, inconsistencies, and inefficiencies. Lack of standardization in medical terminology and coding not only hinders information sharing but also introduces patient safety risks, administrative delays, and claim denials. Thus, the implementation of ontologies is not merely a technical enhancement but a strategic necessity for realizing the full potential of big data analytics in healthcare.

### 1.2. Research Objectives and Scope

The aim of this study is to examine deeply into the foundational concepts of both ontologies and big data, exploring their collaborative application to address complex challenges in healthcare. This analysis will identify key existing challenges, propose how ontology-driven approaches can mitigate these issues, and project future directions for this evolving field. This work encompasses the theoretical underpinnings of ontologies and big data, practical methodologies for their integration, and real-world applications across various clinical domains, including clinical decision support, personalized medicine, and public health initiatives. Furthermore, it will examine the technological toolchains that enable these advanced analytics, including their integration with big data frameworks such as Hadoop and Spark.

To provide a structured analysis, ontology-driven healthcare analytics will be classified into six categories: (i) ontology-driven integration frameworks, (ii) semantic modeling for metadata enrichment, (iii) OBDA, (iv) basic semantic data management, (v) ontology-based reasoning for decision support, and (vi) semantic annotation for unstructured data. Additionally, the work highlights emerging trends such as artificial intelligence, machine learning, the IoT, and real-time analytics, underscoring their role in shaping sustainable, interoperable, and high-performance healthcare ecosystems. The structure of this work, integrating diverse areas such as computer science, data science, and medical domain expertise, underscores the inherently multidisciplinary nature of biomedical informatics. A holistic understanding of ontology and big data analytics in healthcare necessitates synthesizing knowledge from these varied fields to ensure that the analysis is not only technically sound but also clinically relevant and strategically insightful.

The remainder of this paper is structured as follows. Section 2 presents the background analysis, introducing core concepts of ontology, its fundamental principles, development methodologies, and applications in healthcare, along with an overview of Big Data characteristics and challenges. Section 3 describes the research strategy, detailing the research questions, systematic literature search process, and inclusion–exclusion criteria used for study selection. Section 4 proposes the ontology-integrated framework for healthcare data analytics, explaining its layered architecture that integrates ontology with Big Data tools for efficient data ingestion, storage, processing, and decision support. Section 5 discusses ontology-driven approaches for Big Data analytics, highlighting semantic integration, annotation, and ontology-mediated querying techniques. Section 6 illustrates the applications of ontology-driven Big Data analytics in healthcare, covering use cases such as clinical decision support, predictive disease diagnosis, and personalized medicine. Section 7 provides an in-depth discussion and outlines future research directions, focusing on scalability, real-time analytics, and AI-IoT integration. Finally, Section 8 concludes the paper by summarizing key findings, insights, and potential avenues for advancing ontology-driven healthcare analytics.

## 2. Background Analysis

This section introduces core ontology principles, types, languages, development methods, and knowledge-representation models, then examines healthcare-specific ontologies, Big Data characteristics and challenges, comparative literature, and reasoning techniques for semantic inference and decision support. Section 2.1 covers ontology fundamentals. Section 2.2 reviews healthcare ontologies. Section 2.3 discusses big data in healthcare. Section 2.4 analyzes existing studies. Section 2.5 outlines ontology reasoning techniques. Section 2.6 highlights emerging trends and future directions.

### 2.1 Ontology Fundamentals

The foundations of ontologies in healthcare provide the structural backbone for organizing and interpreting complex medical knowledge. This section first clarifies what ontologies are by highlighting their essential components and major classifications. It then examines ontology languages, emphasizing their intended use, advantages, challenges, and roles in healthcare applications. Following this, ontology development methodologies are discussed, including top-down, bottom-up, middle-out, and widely recognized frameworks. The section concludes with an overview of knowledge representation models that play a critical role in healthcare analytics.

#### 2.1.1 Definition and Types of Ontologies

Ontology is the discipline that studies the structure of reality, organizing it into integrative levels such as physical, biological, mental, and cultural. These levels form the basis for more complex domains. In the context of knowledge organization, ontology plays a vital role by providing structured models and guiding frameworks for representing and managing knowledge. The key components of ontologies include classes (concepts), relationships, attributes (properties), and instances. Classes serve as the fundamental categories or entities within a domain, such as Patient, Doctor, or Disease in healthcare. Relationships define how these classes interact, for instance, a Doctor administers Treatment to a Patient. Attributes describe the characteristics associated with classes, such as a Product having a Price or Brand. Instances represent specific real-world examples of classes, such as John Doe as a Patient. Together, these elements establish the foundation for representing and structuring domain knowledge.

Beyond these foundational components, ontologies can be categorized into several types depending on their scope and purpose. Terminology ontologies are designed to standardize vocabulary for describing health concepts, such as SNOMED-CT. Domain-specific ontologies support specialized areas of healthcare, with the Disease Ontology being a notable example. Upper-level ontologies provide high-level, cross-domain frameworks, as illustrated by the Basic Formal Ontology (BFO)[1]. Finally, application ontologies combine both domain and task-specific needs, such as those developed for modeling patient workflows in clinical environments [18].

The hierarchical classification of ontologies, ranging from upper-level to domain-specific and application-specific, reflects the necessity for both broad conceptual coherence and granular detail in knowledge representation. This layered approach allows for effective management of complexity: foundational principles are established at higher levels, while specific nuances and contextual details are captured at lower, more specialized levels [19]. This ensures both the wide applicability of the general framework and the precise representation of domain-specific information [20].

The semantic consistency ensured by ontologies is of paramount importance in healthcare. In this field, even minor variations in terminology or definition can lead to significant consequences, such as misdiagnosis or the misinterpretation of crucial patient data. The absence of a standardized and unambiguous way to define concepts can introduce errors and inefficiencies, directly impacting patient safety and the quality of care [21] [22]. By providing a consistent framework, ontologies mitigate these risks, fostering more accurate data interpretation and reliable decision-making. In this context, the World Wide Web Consortium (W3C)[2] plays a crucial role by developing and maintaining ontology-related standards such as Resource Description Framework (RDF), RDF Schema (RDFS), and Web Ontology Language (OWL), which provide the formal foundations for semantic interoperability across healthcare systems and beyond [23].

2.1.2 Ontology Languages

Ontology languages are formal languages designed to encode and express ontologies. It describes *what knowledge should be represented* rather than *how to compute it*. Most ontology languages are rooted in first-order logic or description logic (DL), enabling automated reasoning.

As shown in **Table 1**, ontology languages range from basic data structuring tools to highly expressive reasoning frameworks and specialized validation or vocabulary systems. Extensible Markup Language (XML) focuses on data serialization but lacks inherent semantic meaning, making it suitable for applications such as HL7 v2 messages [24] [25]. RDF introduces a model for representing resources and their relationships as triples, enabling interoperability and metadata processing important in contexts like EHR metadata exchange. RDFS extends RDF with schema constructs, offering lightweight support for defining classes and properties, often applied in healthcare vocabularies. OWL builds on RDF and RDFS by providing rich semantic representation and reasoning support, which makes it highly valuable for advanced healthcare ontologies such as SNOMED-CT and the BFO. Other languages play complementary roles: DAML+OIL[3], a precursor to OWL, contributed to early medical ontology prototypes; Shapes Constraint Language (SHACL) and SPARQL Inferencing Notation (SPIN) ensure RDF data validation and rule-based reasoning in healthcare workflows; and Simple Knowledge Organization System (SKOS) supports controlled vocabularies and disease classifications [26] [27].

[1] https://basic-formal-ontology.org/
[2] https://www.w3.org/
[3] https://www.w3.org/TR/daml+oil-reference/

**Table 1.** Comparison of Ontology Languages

| Language | Purpose | Strengths | Limitations | Healthcare Example |
|---|---|---|---|---|
| XML | Data structuring & serialization | Flexible, widely adopted | No semantic meaning | HL7 v2 messages |
| RDF | Represent resources and relationships as triples | Enables interoperability & metadata processing | Limited expressiveness | EHR metadata exchange |
| RDFS | Extend RDF with schema (classes, subclasses, properties) | Lightweight schema definition; easy to use | Limited expressiveness; weaker than OWL | RDF schema for EHR vocabularies |
| OWL | Rich semantic representation with reasoning support | Formal logic, supports AI-driven reasoning | Computationally complex | SNOMED-CT, BFO |
| DAML+OIL | Early ontology language for richer semantics than RDF | Introduced restrictions, class hierarchies | Deprecated, replaced by OWL | Early medical ontology prototypes |
| SHACL | Validate RDF data against constraints | Ensures data integrity; supports RDF validation | Not designed for reasoning | Validation of FHIR RDF data |
| SPIN | Rule-based reasoning using SPARQL | Flexible rule-based inference | Less standardized, heavy for large datasets | Workflow validation in clinical RDF |
| SKOS | Support vocabularies, taxonomies, thesauri | Lightweight, interoperable | Limited reasoning support | Disease classification vocabularies |

The use of DL as the foundational basis for ontology languages like OWL is critically important. DLs are formal knowledge representation languages that provide the necessary rigor for automated reasoning and consistency checking. In high-stakes applications such as healthcare, where accuracy and reliability are paramount, this formal foundation ensures that the knowledge embedded within ontologies can be processed and reasoned upon by machines with a high degree of confidence. This capability is non-negotiable for supporting clinical decision-making and ensuring patient safety.

2.1.3 Ontology Development Methodologies

Ontology development is a systematic process of creating, organizing, and refining domain concepts and their interrelationships. The choice of methodology depends on the maturity of domain knowledge and the characteristics of available data. Figure 1 illustrates the general phases of ontology development, highlighting the iterative and structured nature of the process. Ontology development begins with determining the scope, defining the purpose, domain boundaries, and objectives. Next, reuse is considered by identifying existing ontologies or knowledge bases to adapt. The taxonomy is then defined, establishing a hierarchical structure of classes and subclasses. Developers enumerate terms, listing all relevant concepts and instances, and define constraints, specifying rules and logical conditions. Properties are defined to model relationships and attributes, and finally, instances are created to populate the ontology with real-world examples [28] [29].

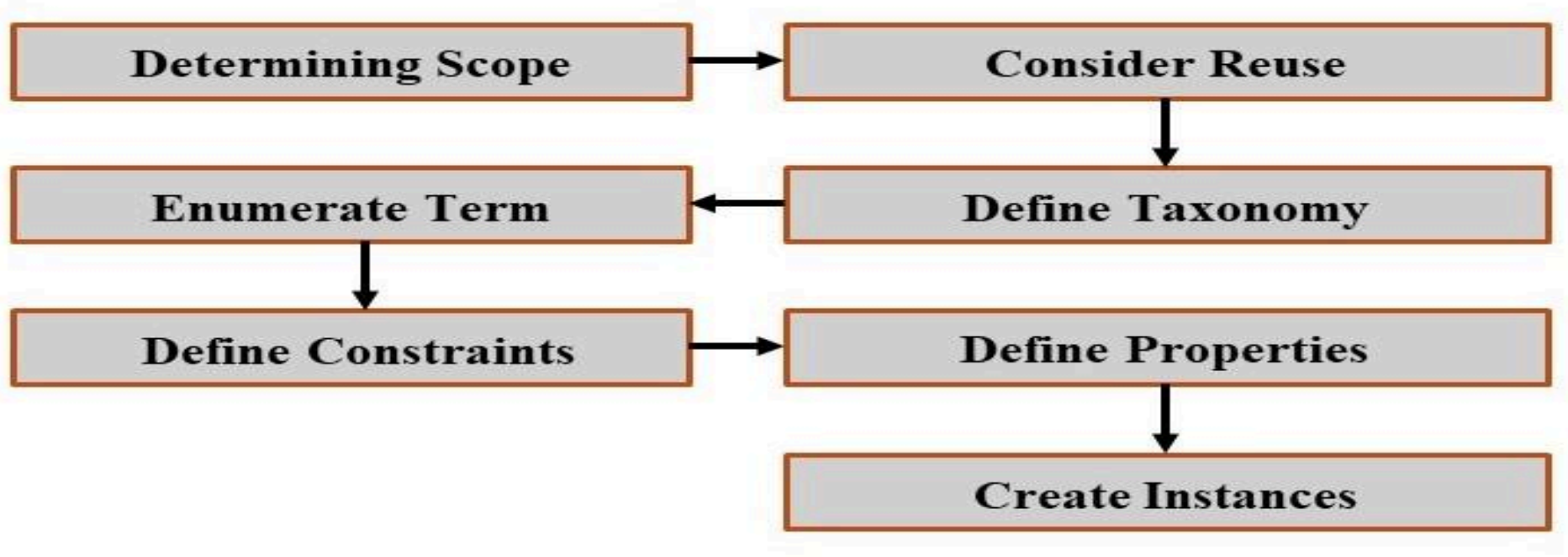

Figure 1. Ontology development phases

As summarized in Table 2, the most common approaches are Top-Down, Bottom-Up, and Middle-Out.

Table 2. Comparison of Ontology Development Approaches

| Approach | Explanation | Steps | Advantages | Disadvantages | Use Case |
|---|---|---|---|---|---|
| **Top-Down** | Start with broad, abstract concepts, refine into specifics | Define domain → Identify abstract concepts → Refine subclasses → Define relationships | Clear hierarchy, domain-centric, comprehensive | May overlook details, time-consuming | Healthcare, Finance |
| **Bottom-Up** | Start with detailed data, group into abstract classes | Collect datasets → Identify entities → Group into classes → Build hierarchies | Captures nuanced details, scalable with new data | May lack overall structure, risk of disorganization | Retail, Logistics |
| **Middle-Out** | Begin with key mid-level concepts, expand upward & downward | Identify mid-level concepts → Generalize upward → Refine downward | Balanced structure, bridges abstraction & detail | Requires deep domain knowledge, risk if mid-level is misdefined | IT services, Project Management |

The selection of an ontology development methodology is a strategic decision profoundly influenced by the maturity of the domain understanding and the intrinsic nature of the available data. This choice directly impacts the ontology's eventual structure, its completeness, and its long-term maintainability. An inappropriate methodological selection can lead to an ill-suited or difficult-to-manage ontology, underscoring the critical importance of this initial decision [30].

Considering the inherent complexity and heterogeneity of healthcare data, which often includes fragmented, unstructured, and diverse information , a purely top-down or bottom-up approach may prove insufficient. A top-down strategy might miss crucial nuances embedded in granular data, while a bottom-up approach could struggle to achieve a coherent and overarching high-level structure. Consequently, a hybrid approach, such as the Middle-Out methodology, or an iterative combination of top-down and bottom-up strategies, is frequently necessary [31]. This provides the flexibility required to reconcile the need for broad conceptual models with the granular realities of clinical data, allowing for continuous refinement and adaptation in a dynamic domain.

Popular ontology development methodologies offer structured approaches to guide the creation process. Methontology is a well-defined, iterative methodology that emphasizes systematic knowledge acquisition, documentation, and lifecycle management. In contrast, Uschold and King's methodology adopts a goal-driven perspective, where clearly defining the purpose and scope serves as the foundation for ontology design. Meanwhile, the NeOn methodology provides a collaborative framework tailored for building and managing networked ontologies, with strong support for ontology reuse, integration, and evolution in dynamic environments [32].

### 2.1.4 Knowledge Representation Models in Healthcare Analytics

Knowledge representation (KR) models form the backbone of healthcare analytics by structuring complex medical data into computable formats that support reasoning and decision-making. Ontology-based models (like., SNOMED CT, ICD-10, LOINC) enable semantic interoperability and ontology-driven querying for clinical decision support. Semantic networks capture relationships between medical entities as graphs, supporting disease association and comorbidity analysis. Frame-based models represent healthcare concepts as attribute–value pairs, facilitating case-based reasoning. Rule-based models (like., SWRL, SPIN) encode clinical guidelines and generate automated alerts [33]. To address uncertainty, probabilistic models such as Bayesian networks and fuzzy logic frameworks are used for risk prediction and diagnostic reasoning. Emerging hybrid models integrate ontologies with probabilistic and rule-based reasoning, enabling more advanced applications like personalized medicine and real-time monitoring [34]. Collectively, these KR models provide the foundation for knowledge-driven healthcare analytics, as illustrated in *Figure 2*, which categorizes the major models applied in clinical and public health contexts.

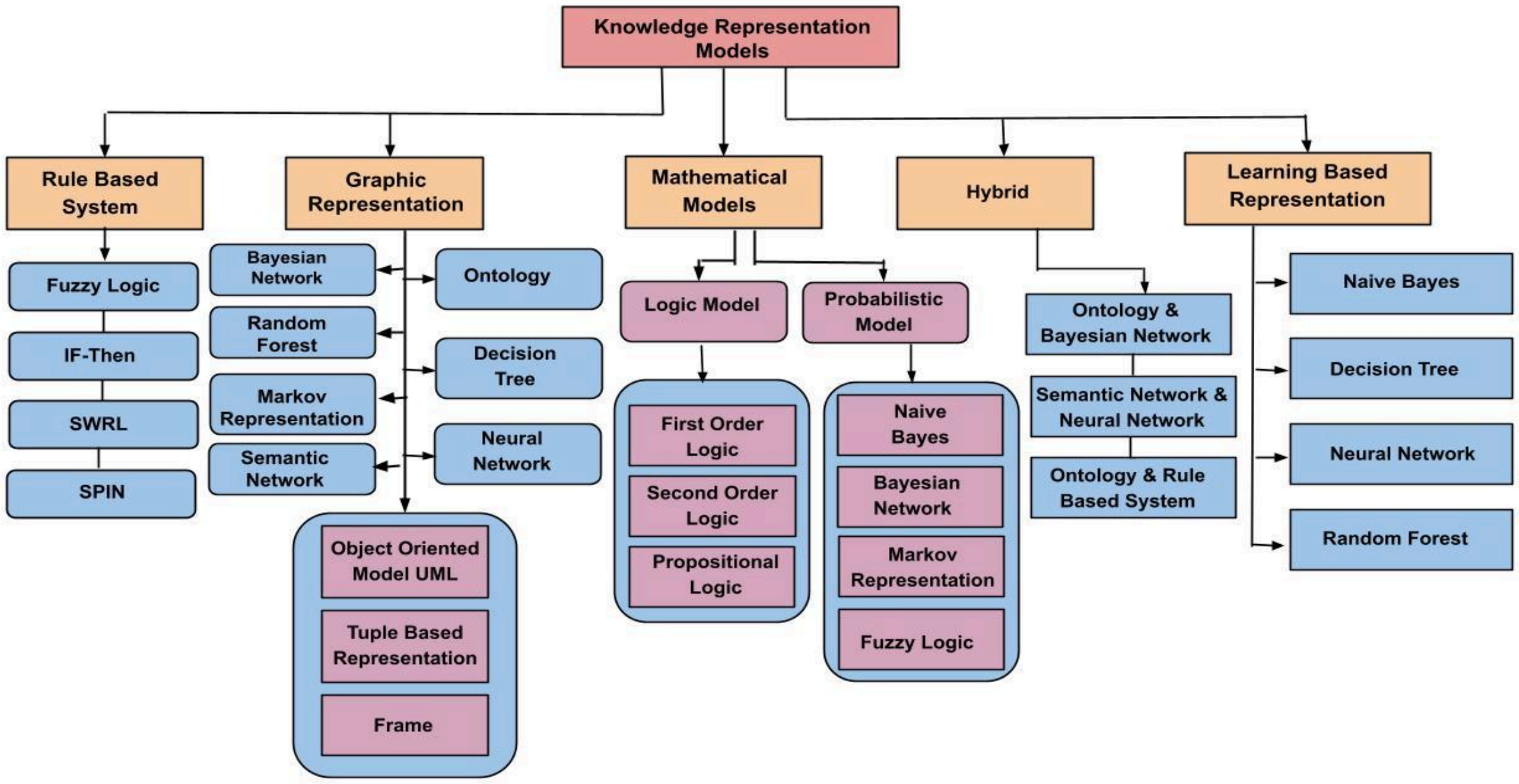

Figure 2. Categories of Knowledge Representation Approaches

## 2.2 Healthcare Ontologies

Healthcare ontologies provide structured and standardized representations of medical knowledge, enabling semantic interoperability, data integration, and advanced analytics across diverse healthcare systems. These ontologies define core concepts, relationships, and properties relevant to clinical care, laboratory testing, genomics, and biomedical research. Widely used healthcare ontologies include SNOMED CT for clinical terminology, LOINC for laboratory measurements, ICD for disease classification, UMLS for integrating multiple biomedical vocabularies, MeSH for biomedical literature indexing, and domain-specific ontologies such as Gene Ontology (GO), Disease Ontology (DOID)[4], Human Phenotype Ontology (HPO), RxNorm[5], RadLex, and Foundational Model of Anatomy (FMA ) [35]. These ontologies serve as the semantic backbone of healthcare data systems, supporting precise communication, knowledge sharing, and interoperability across clinical, research, and public health applications. Table 3 summarizes the key healthcare ontologies, their domains, purposes, and typical use cases.

**Table 3.** Existing healthcare ontology

| Category | Ontology | Domain / Scope | Purpose |
|---|---|---|---|

[4] https://disease-ontology.org/
[5] https://www.nlm.nih.gov/research/umls/rxnorm/overview.html

| | | | |
|---|---|---|---|
| Clinical & EHR Ontologies | SNOMED CT | Clinical healthcare concepts (diseases, findings, procedures, body structures) | Standardized terminology for recording & sharing data |
| | ICD (10/11) | Diseases, disorders, injuries, health conditions | Classification for epidemiology, billing |
| | UMLS | Meta-thesaurus linking multiple vocabularies | Integrates vocabularies, cross-mapping |
| | HL7 FHIR Ontology | Healthcare data exchange | Defines resources & semantic models |
| Laboratory & Imaging Ontologies | LOINC | Lab tests, clinical measurements, observations | Standardizes lab test names & results |
| | RadLex | Radiology terminology | Standardized radiology vocabulary |
| Biomedical Research Ontologies | MeSH | Biomedical literature | Controlled vocabulary for indexing |
| | GO | Genomics, molecular biology | Representation of gene function |
| | DOID | Human diseases | Standardized disease classification |
| | HPO | Phenotypic abnormalities | Describes traits in rare diseases |
| | RxNorm | Clinical drugs & medications | Normalized drug names, drug relations |
| | FMA | Human anatomy | Anatomy ontology |

### 2.2.1 Ontology Development for Specific Healthcare Domains

Developing ontologies for particular healthcare domains is crucial to ensure semantic consistency, interoperability, and advanced analytics customized to the unique requirements of each medical specialty. Unlike generic ontologies such as SNOMED CT, ICD, or LOINC, which provide a broad framework for representing healthcare data, domain-specific ontologies focus on capturing specialized knowledge with high granularity. For example, cardiology ontologies define detailed concepts around heart diseases, diagnostic tests, and treatment pathways, while oncology ontologies capture tumor classifications, cancer biomarkers, and therapy protocols. Similarly, neurology ontologies represent neurological disorders, brain structures, and cognitive functions, whereas pharmacology and drug ontologies model medications, dosages, and interactions for safer and more effective treatment planning [36].

Public health ontologies, on the other hand, support disease surveillance, outbreak detection, and policy-making by combining demographic, epidemiological, and clinical datasets. Mental health ontologies are increasingly important as they capture psychological disorders, behavioral traits, and therapy interventions, often integrating data from wearable devices for stress and mood monitoring [37]. Collectively, these domain-specific ontologies not only facilitate knowledge sharing and system interoperability but also empower advanced clinical decision support systems, predictive analytics, and research innovations in their respective fields. Table 4 for examples of disease-specific ontologies.

Table 4. Disease specific ontology

| Specialty | Ontology | Focus | Purpose |
|---|---|---|---|
| Neurological Disorders [38] | Alzheimer's Disease Ontology (ADO) | Pathology, biomarkers, genetics, progression | Standardize Alzheimer's research |
| | Parkinson's Disease Ontology (PDON) | Motor & non-motor symptoms, stages, treatments | Support neurodegeneration research |
| Metabolic & Endocrine Disorders [26] | Diabetes Mellitus Ontology (DMO) | Types, complications, treatments | Structured knowledge for diabetes |
| | Obesity Ontology | Risk factors, comorbidities, genetics | Standardize obesity knowledge |
| Respiratory Diseases [39] | Asthma & COPD Ontology | Symptoms, triggers, therapies | Semantic interoperability |
| Cancer & Oncology [40] | Breast Cancer Ontology (BCO) | Tumor classification, biomarkers, treatments | Standardize oncology research |
| Infectious Diseases [41] | IDO (HIV, Malaria, COVID-19, etc.) | Pathogen traits, disease progression | Framework for infectious disease |
| | Tuberculosis Ontology (TBO) | TB infection, drug resistance, treatments | TB control & research |
| | Hepatitis Ontology | Hepatitis virus types (A, B, C) | Model hepatitis knowledge |
| Cardio & Stroke [42] | Stroke Ontology | Risk factors, ischemic/hemorrhagic types, treatments | Structured stroke research |

### 2.2.2 Ontology Evaluation and Validation Methods

Ontology evaluation and validation are essential steps in ensuring that a developed ontology is both semantically accurate and practically useful. These processes verify that the ontology meets the intended domain requirements, maintains logical consistency, and performs effectively when applied in real-world systems. Different evaluation methods focus on various aspects of the ontology, including syntactic correctness, logical coherence, structural quality, domain accuracy, query response capability, and usability in applications [43] [44].

Table 5 summarizes common ontology evaluation and validation methods, highlighting their focus, purpose, and the tools or techniques commonly used. For example, lexical and syntactic checks ensure that the ontology is machine-readable and compliant with OWL/RDF standards, while logical consistency validation identifies contradictions and unsatisfiable classes. Structural evaluation assesses taxonomy completeness and connectivity, and domain expert validation confirms that the ontology accurately represents domain knowledge. Competency questions test the ontology's ability to answer relevant queries, while gold standard comparisons measure coverage and accuracy against reference datasets. Application-based evaluation verifies usability and interoperability in practical deployments, and quantitative metrics provide measurable benchmarks for coverage, cohesion, and scalability [45][46].

**Table 5.** Methods for Ontology Quality Assessment and Validation in Healthcare Analytics

| Method | Focus | Purpose | Tools/Techniques |
|---|---|---|---|
| Lexical & Syntactic | Syntax, language compliance | Ensure machine readability & OWL/RDF validity | OWL validators, Protégé checker |
| Logical Consistency | Contradictions, unsatisfiable classes | Guarantee internal logical soundness | Pellet, HermiT, FaCT++ |
| Structural Evaluation | Hierarchy, taxonomy, redundancy | Ensure balanced structure & connectivity | OntoMetrics, Protégé plugins |
| Domain Expert Validation | Accuracy of domain knowledge | Confirm semantic relevance & correctness | Expert review, Delphi method |
| Competency Questions | Query response accuracy | Test if ontology answers domain-specific queries | SPARQL, reasoning tasks |
| Gold Standard Comparison | Match with reference ontology/dataset | Check coverage, accuracy, completeness | Mapping tools, similarity metrics |
| Application-Based | Real-world system performance | Validate usability, interoperability | Case studies, pilot integration |
| Quantitative Metrics | Coverage, cohesion, scalability | Provide measurable quality benchmarks | Coverage ratio, cohesion metrics |

## 2.3 Big Data Analytics in Healthcare

Big Data has emerged as a transformative force in healthcare, integrating diverse datasets for advanced analytics. This section highlights key types of healthcare data, outlines the core characteristics of Big Data, and examines major challenges, emphasizing the role of ontologies in addressing issues of interoperability, governance, and knowledge discovery.

### 2.3.1 Types of Healthcare Data

The healthcare industry generates an immense and diverse array of data, having undergone a significant transformation from predominantly paper-based records to extensively digitized information. This data deluge encompasses various forms, each contributing uniquely to the comprehensive patient profile and broader health insights [47] [48].

Key types of healthcare data include:

- **EHRs and Electronic Medical Records (EMRs):** These form the bedrock of patient information, originating from doctor visits and encompassing a patient's medical history, physical examination findings, treatment outcomes, current health conditions, and overall outcomes data. EHRs are particularly complex as they can exist in structured, semi-structured, or entirely unstructured formats.
- **Imaging Studies:** This category comprises unstructured data such as MRIs, CT scans, X-rays, and PET scans. These visual data represent the fastest-growing segment of healthcare data, presenting unique challenges for storage and analysis due to their size and complexity.
- **Genomics and 'Omics Data:** This refers to high-throughput data derived from advanced analyses like genomics, proteomics, metabolomics, pharmacogenomics, and disease omics. These datasets provide profound insights into the complex biochemical and regulatory processes within living organisms. A significant characteristic of 'omics data is their inherent heterogeneity, often being stored in disparate data formats.
- **Claims Data:** These consist of electronic financial transactions related to health insurance claims. This data is frequently utilized for research and various analytical purposes, offering a different lens on healthcare utilization and costs.
- **Wearable Devices and Remote Monitoring:** A rapidly expanding source, these devices generate real-time data on a multitude of physiological parameters, including physical activity, heart rate, sleep patterns, and blood pressure.[2] This continuous stream of information is invaluable for early detection of potential health issues and ongoing patient monitoring.
- **Patient Surveys and Lifestyle Information:** This data encompasses patient experiences, behavioral patterns, lifestyle choices (like, tobacco use, exercise habits), and crucial social determinants of health. Such information augments formal medical data, providing a more holistic view of patient well-being.
- **Administrative Data:** This includes factual information about health insurance, such as eligibility and membership details, as well as unique identifiers for providers and facts about the nature of healthcare institutions.
- **Clinical Notes:** Often captured as unstructured natural language text, clinical notes contain rich, detailed patient information that poses significant challenges for automated analysis due to their free-form nature.

The inherent heterogeneity and fragmentation of healthcare data across multiple types and sources present a major technical challenge in achieving a holistic view of patient health and population-level trends. Patient information is often siloed across hospitals, clinics, laboratories, insurers, and personal health devices, making it difficult to construct a complete and unified record for analysis. To address this, healthcare data must be systematically collected from diverse origins such as medical records, monitoring devices, administrative systems, and patient surveys, and then organized into a structured flow [49]. As illustrated in Figure 3, the collected data is categorized by its focus—patient-level data for clinical insights, provider-level data for operational efficiency, and policy-level data for systemic evaluation. Patient-level data supports tasks like disease prediction and comorbidity analysis; provider-level data improves physician collaboration and hospital coordination; and policy-level data informs cost management and performance monitoring. This structured integration process demonstrates how fragmented datasets, when properly managed, can be transformed into actionable insights that advance both individual patient care and broader healthcare decision-making.

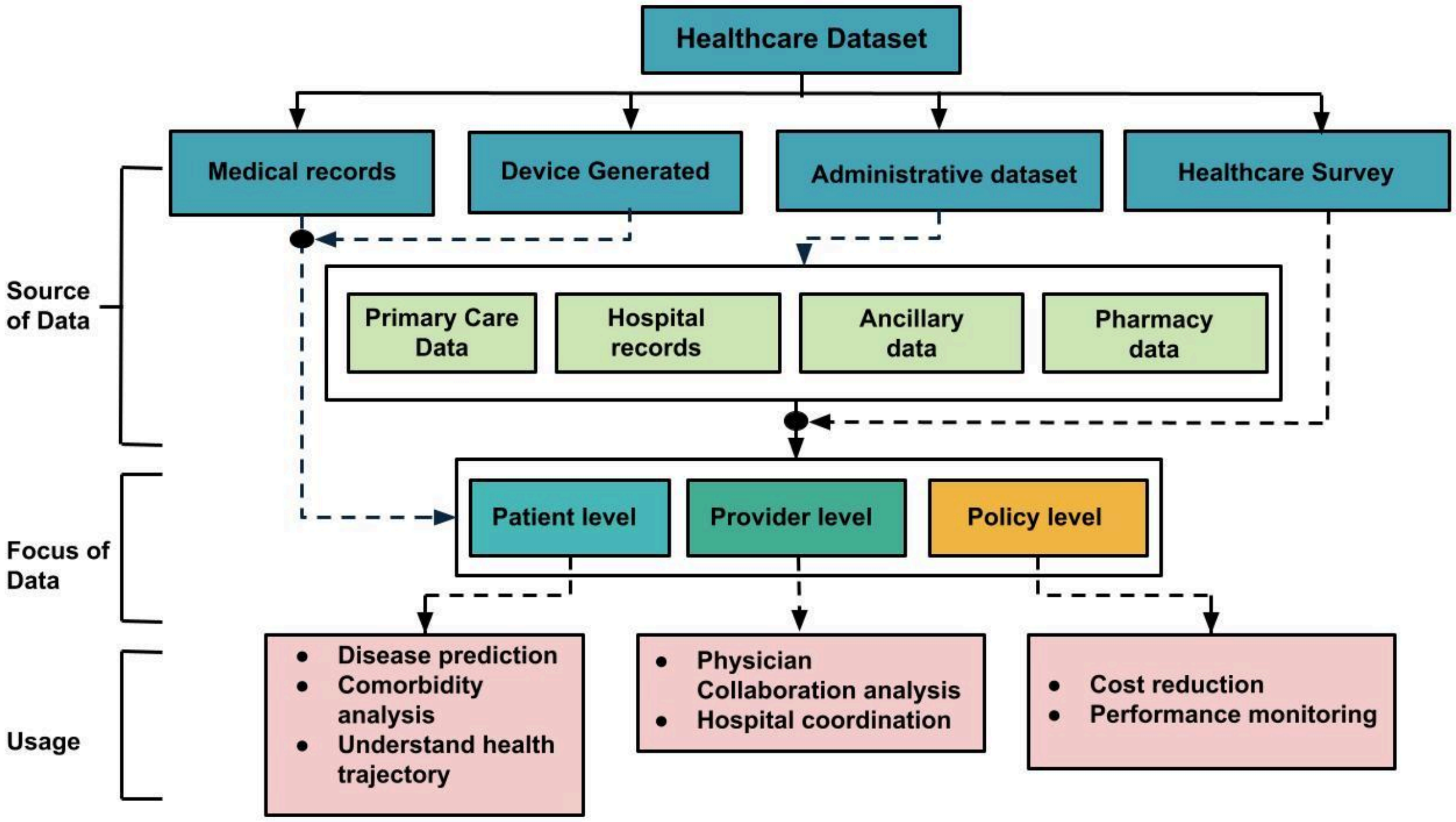

Figure 3. A conceptual model showing the flow of healthcare datasets from their diverse sources to their specific analytical applications.

### 2.3.2 Big Data Characteristics

Big Data is typically defined by a set of characteristics, commonly referred to as the "V's of Big Data" shown in Figure 4. These characteristics describe not only the scale and complexity of healthcare datasets but also the core challenges that necessitate the adoption of Big Data Analytics [47]. Addressing these challenges is essential to unlock the value proposition that Big Data offers in healthcare [50].

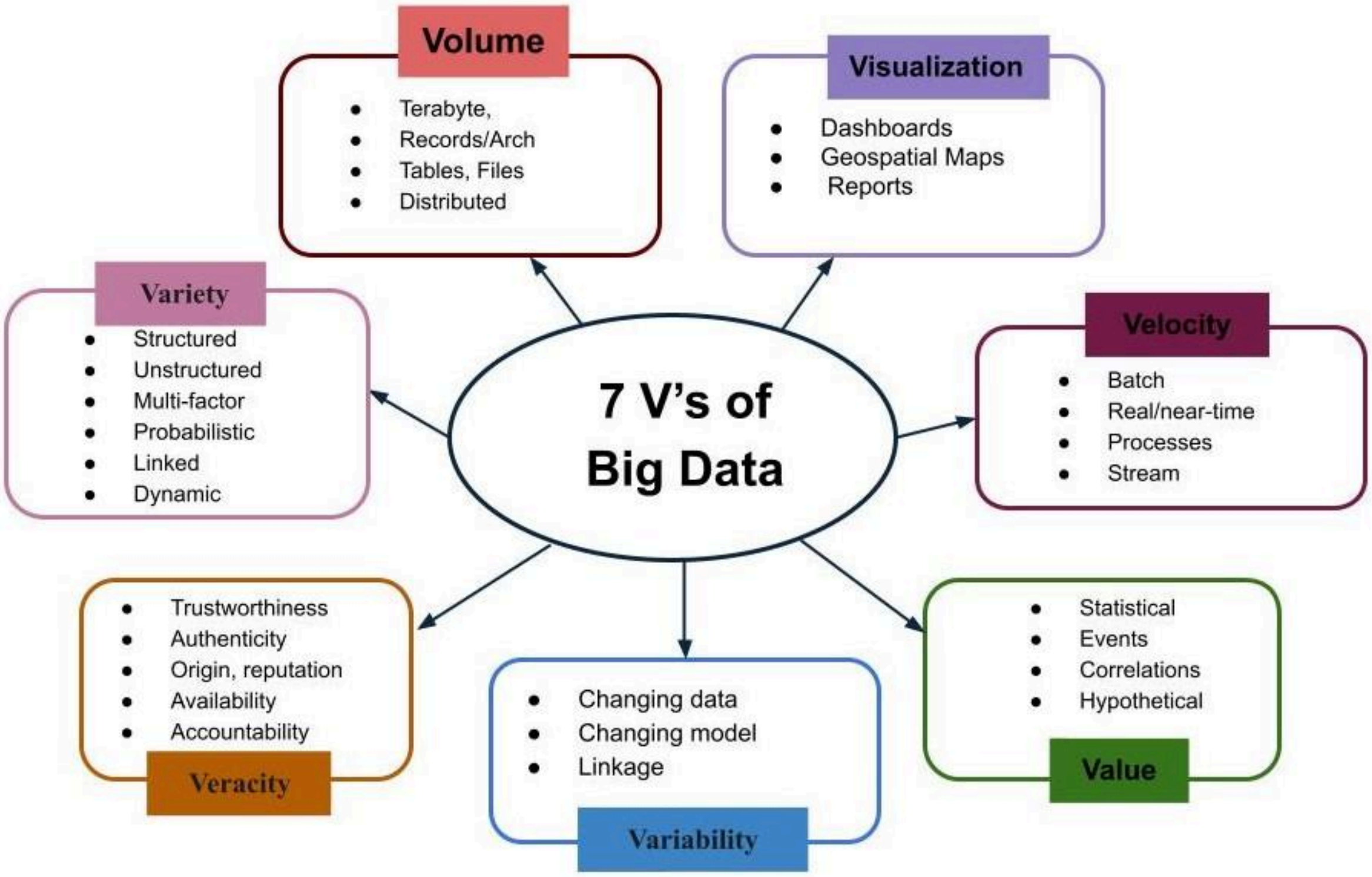

**Figure 4.** The 7 V's of Big Data

- **Volume**: Refers to the immense quantity of data generated and stored. In healthcare, this includes billions of electronic health record (EHR) entries, vast archives of medical imaging, and petabytes of genomic sequences that continue to grow exponentially.
- **Velocity**: Denotes the speed at which data is generated, collected, and processed. Examples include real-time patient monitoring in intensive care units (ICUs), continuous data streams from wearable devices, and rapid diagnostic test results.
- **Variety**: Highlights the diversity of data formats, ranging from structured (lab test results), semi-structured (JSON, XML), and unstructured data (clinical notes, pathology slides, genomic data). Sources are equally diverse, including hospitals, laboratories, research centers, and IoT devices.
- **Veracity**: Represents the trustworthiness, accuracy, and quality of data. Healthcare data often suffers from incompleteness (missing entries), inconsistency (mismatched sources), or errors, all of which can significantly affect clinical outcomes.
- **Value**: Captures the ultimate goal of Big Data — generating actionable insights that improve patient outcomes, enable disease prediction, optimize treatments, and reduce healthcare costs.
- **Variability**: Refers to the inconsistency in data flows and changing structures over time. Healthcare data can fluctuate in frequency, precision, and quality, adding complexity to processing and analysis. For instance, data from wearable devices can vary widely across patients and conditions.
- **Visualization**: Emphasizes the need for effective presentation of analytical outcomes.

Dashboards, charts, geospatial maps, and interactive reports help clinicians and policymakers interpret data patterns, track disease spread, and make informed decisions.

These characteristics collectively define the problem space that Big Data Analytics seeks to address in healthcare. Advanced distributed processing platforms (e.g., Hadoop, Spark) and sophisticated analytical techniques are essential to manage data with high volume, velocity, variety, and variability, while ensuring veracity. Successfully addressing these challenges is what ultimately enables healthcare systems to derive value [51] [52].

Among all dimensions, veracity holds unique importance in healthcare. Unlike other industries, errors in medical data directly affect patient safety and clinical decision-making. Inaccurate records or inconsistent terminology can lead to misdiagnoses, inappropriate treatments, or adverse drug interactions. Thus, ensuring data integrity, accuracy, and security is not merely operational but a life-saving imperative [47].

2.3.3 Challenges in Healthcare Big Data Analytics

As shown in Figure 5, healthcare big data analytics faces multiple challenges such as heterogeneity of data sources, data quality, privacy, scalability, and semantic consistency [53]. Healthcare data comes from diverse sources such as EHRs, medical imaging, lab reports, wearable devices, and genomic datasets, often in different formats and standards. This heterogeneity of data sources poses a significant challenge for data integration and interoperability, as combining information across systems while maintaining semantic consistency is difficult. Ontologies provide a solution by offering a common conceptual framework that maps diverse data sources into a unified semantic model. It enables semantic mapping between different schemas, allowing systems to exchange information meaningfully and improving interoperability across heterogeneous sources [54] [55] [56] [57].

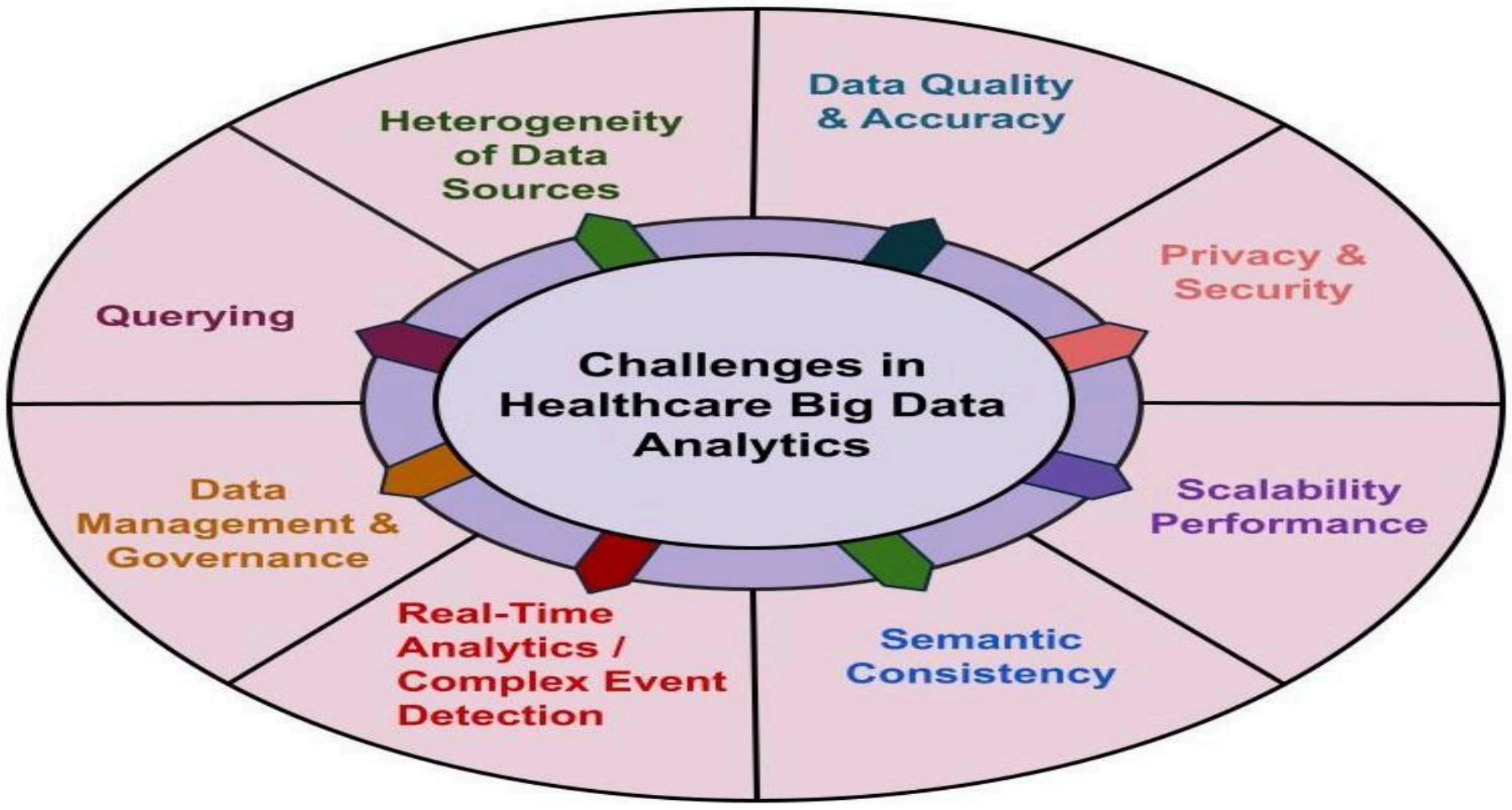

Figure 5. Challenges in Healthcare Big Data Analytics

In addition, querying and knowledge discovery over large, complex datasets is challenging due to the diversity and volume of data. Ontologies support semantic querying and reasoning [58], allowing advanced queries that can infer relationships not explicitly stored in the data. For example, ontology-based reasoning can identify patients at risk of certain conditions by correlating multiple data points across sources. Data management and governance is another challenge. Given the sensitivity of healthcare data, it is important to structure metadata, define rules for data lifecycle management, quality control, and ensure regulatory compliance. Ontologies enhance governance by tracking data provenance, access policies, and usage, which is critical in healthcare settings [59].

Data quality and accuracy are also major concerns, as inconsistent or incomplete data can reduce the reliability of analytics. Ontologies help by enforcing semantic constraints and validation rules, such as standardizing units of measurement or ensuring biologically plausible values, thus improving the integrity of data [60]. Privacy and security are equally important, since healthcare data contains highly sensitive patient information. Ontologies contribute by defining sensitivity levels and access control policies for patient data, enabling automated anonymization or controlled sharing while maintaining semantic clarity. [61]

Real-time analytics and complex event detection, such as monitoring intensive care unit patients, are challenging due to high data velocity and volume. Ontology-driven semantic frameworks facilitate event correlation and interpretation, allowing real-time detection of critical conditions from multiple data streams [62]. Scalability and performance in big data environments present another challenge. Ontologies address this by structuring information hierarchically and enabling semantic indexing, which optimizes queries and reduces processing bottlenecks [63][64]. Finally, ensuring semantic consistency and standardization is crucial so that concepts like “blood pressure” or “myocardial infarction” are interpreted consistently across systems. Without this, analytics and interoperability efforts would be undermined.

### 2.4 Comparative Analysis of Existing Studies

Ontology-driven healthcare studies shown in Table 6 cover diverse domains including IoT-based monitoring, EHR interoperability, cardiovascular disease representation, pregnancy and diabetes data preparation, COVID-19 remote monitoring, and healthcare security. They utilize semantic technologies such as OWL, RDF/SPARQL, SWRL, HL7-FHIR, SNOMED CT, OMOP2OBO, and application-specific ontologies, integrated with platforms like Hadoop, Spark, Kafka, NoSQL databases, graph models, and blockchain. Key contributions include enhanced semantic interoperability, improved data organization, faster query performance, decision support, anomaly detection, and multi-center data integration. Despite these strengths, most works are restricted to disease-specific contexts, encounter scalability issues under large datasets, and remain conceptual or proof-of-concept without real-time deployment in production healthcare systems, leaving a gap in demonstrating ontology performance in large-scale environments.

**Table 6. Comparative Analysis**

| # | Author & Year | Domain Focus | Big Data Platform / Storage | Ontology / Semantic Tech Used | Ontology Type | Contribution | Key Limitations |
|---|---|---|---|---|---|---|---|
| [65] | Shah et al., 2015 | Medical & Oral Health | Not specified | OWL 2, SWRL, Protégé, Pellet | Cross-domain (OSHCO) | Oral–systemic health ontology; reasoning & | Complex modeling, expert dependent, |
| | | | | | | decision support | reasoning overhead |
| [66] | Cui & Zhang, 2016 | Ontology QA (SNOMED CT, FMA, GO) | Hadoop, MapReduce | MaPLE (lattice-based evaluation) | Structural auditing ontology | Scalable QA; reduced runtime (months → hours) | Focused on structure; semantics less covered; resource heavy |
| [67] | Ullah et al., 2017 | Healthcare IoT Interoperability | Cloud IoT, RDF Triple Stores | SIMB-IoT, RDF, SPARQL | Domain (IoT healthcare) | RDF-based semantic interoperability; drug recommendation | Only semantic layer; syntactic/security limits; small-scale |
| [68] | Mezghani et al., 2017 | Healthcare IoT / Cognitive Systems | Kafka, Storm, HDFS, Spark, Fuseki | Cognitive patterns, WH_O ontology, SPARQL | Domain (wearable healthcare) + cognitive | Model-driven methodology; cognitive diabetes monitoring system | High complexity; ontology overhead; scalability tied to IT infrastructure |
| [69] | Liyanage et al., 2018 | Routine Clinical Data (Pregnancy) | Clinical DBs (RCGP RSC, CMRs) | OWL, Protégé, BioPortal, SQL | Domain (pregnancy, biomedical) | 3-step ontology process (ontology, coding, query) for transparent case identification | Mapping uncertainty; coding errors; dependent on data quality |
| [70] | Irfan et al., 2019 | Biomedical Text Mining | Not specified | Ontology learning (linguistic, statistical, semantic) | Domain (biomedical) | Survey of ontology learning methods for healthcare text | Expert dependent; semi-automatic; low scalability |
| [71] | Li et al., 2021 | Bridge SHM (Structural Health Monitoring) | Hadoop (HDFS, HBase, Spark) | BSHM ontology (extends SSN, SOSA, QUDT); OWL, RDF, SPARQL, SWRL, R2RML | Domain (bridge SHM) | Fine-grained SHM modeling; anomaly detection; decision support; deployed in big data platform | Focused on girder bridges; limited sensor failure handling; complex modeling |
| [72] | Gupta & Singh, 2021 | IoT / Elderly Healthcare | IoT + Edge Processing | Ontology-based IoT Healthcare System (IHS) | Domain-specific | Improved data organization; faster query response for senior care | Limited evaluation of predictive analytics |
| [73] | Sen & Mukherjee, 2024 | Primary Healthcare / Data Storage | NoSQL (MongoDB) | Ontology-driven schema for semi/unstructured health data | Application Ontology | Optimized schema & query performance for heterogeneous health data | Limited real-world deployment; not real-time |
| [74] | Das & Hussey, 2023 | EHR / Interoperability | HL7-FHIR APIs + Knowledge Graph | ContSys Formal Ontology + HL7-FHIR | Interoperability Ontology | Enabled continuity of care; interoperable data exchange | Limited scalability evaluation under big data |
| [75] | Sabir et al., 2025 | Cardiovascular Disease | RDF Store + SPARQL endpoints | Heart Disease Ontology (HDO) based on SNOMED CT, ICD-10, FHIR | Domain-specific | Detailed cardiovascular ontology; validated via SPARQL queries | Focused on ontology creation; not big data |
| [76] | Matulevičius et al., 2022 | Healthcare Application Security | Blockchain + Semantic Reasoning | HealthOnt Ontology + Blockchain | Security Ontology | Modeled security threats in healthcare blockchain applications | Not focused on big data scalability |
| [77] | Callahan et al., 2023 | Translational Research / EHR | OMOP CDM + OBO Tools | OMOP2OBO mapping (OMOP vocab → OBO Ontologies) | Upper-level + Domain | Unified EHR vocabularies; improved rare disease phenotyping | Manual curation still required |
| [78] | Shahzad et al., 2021 | IoT-based Smart Healthcare / IoHT | Semantic Middleware + Cloud | OWL, Protégé, SPARQL, HermiT Reasoner | Domain-specific + Application | Integrated semantic framework; use cases: Arrhythmia, Prostate Cancer, Leukemia | No physical infrastructure; scalability untested |
| [79] | Sharma et al., 2021 | Remote Patient Monitoring (COVID-19) | IoT Wearables + Cloud Analytics | Ontology-based IoT framework with ECG, PPG sensors | Domain-specific (COVID-19) | Alarm-enabled monitoring system (96.33% accuracy) for COVID-19 | Focused only on COVID-19; device heterogeneity |
| [80] | Balakrishnan et al., 2025 | Multi-center Healthcare Interoperability | Graph DBs + ML (Spark pipelines) | Ontology + Knowledge Graph (SNOMED CT, UMLS, ICD) | Interoperability + Domain | Semantic interoperability framework with ML-driven ontology mapping | Proof-of-concept; privacy/security challenges |
| [81] | Croce et al., 2024 | Healthcare Data Preparation (Diabetes) | Ontology-Based Data Management (OBDM) | OBDM framework for EMR integration | Domain-specific (Diabetes) | Unified 13 years of diabetes EMR data; improved quality for AI analytics | Context-specific (diabetes only); EMR heterogeneity issues |

### 2.5 Ontology Reasoning Techniques

Ontology reasoning enables healthcare systems to go beyond static data representation by supporting knowledge inference, decision support, and semantic data integration. Among the different reasoning paradigms, description logic, rule-based approaches, query-driven methods, and probabilistic/fuzzy reasoning are the most prominent.

#### 2.5.1 Description Logic Reasoning

Description Logic (DL) Reasoning enables formal reasoning over ontologies by using logic-based inference to ensure data consistency and derive implicit knowledge. In healthcare, DL reasoning can automatically check for contradictions in patient records or ontology definitions (consistency checking) and organize concepts hierarchically (classification), such as grouping diseases by type or severity [82]. This supports accurate clinical decision-making, ontology validation, and semantic integration of heterogeneous healthcare data. Tools commonly used include Pellet, HermiT, Fact++, and OWL API [83]. Challenges include computational complexity for large-scale ontologies and handling evolving medical knowledge efficiently.

#### 2.5.2 Rule-Based Reasoning

Rule-based reasoning models *if–then* logic in ontologies using SWRL, SPIN, or Jena rule engines, enabling clinical guidelines and care pathways to be encoded into healthcare systems. For example, a rule may infer obesity-related risks if a patient has a BMI > 30 and a family history of Type 2 diabetes. In oncology, such rules support tumor classification by combining genetic and imaging data. This approach is flexible, human-readable, and integrates medical knowledge into analytics, with tools like Protégé SWRLTab, Jena, and Drools. However, it faces challenges in scalability with large datasets and managing conflicting rules [84].

### 2.5.3 Query-Based Reasoning

Query-based reasoning leverages SPARQL, SPARQL-DL, or GraphQL extensions to perform semantic retrieval and reasoning over RDF-based healthcare data, combining explicit ontology knowledge with inferred facts. In healthcare, it can identify patients with specific conditions and treatments, such as those with hypertension on beta-blockers who have abnormal kidney function, supporting drug-safety monitoring. At a population level, it enables pattern extraction for epidemiology surveillance, e.g., tracking outbreaks using EHR and IoT data. Key benefits include dynamic data retrieval, scalability via federated SPARQL endpoints, and integration across distributed healthcare sources [85]. Common tools include Virtuoso, GraphDB, Stardog, and Blazegraph, while challenges involve query optimization and high computational overhead for complex joins [86].

2.5.4 Probabilistic and Fuzzy Reasoning

Probabilistic and Fuzzy Reasoning address uncertainty, vagueness, and incomplete evidence in healthcare data by extending ontologies beyond crisp logic. Probabilistic reasoning assigns likelihoods to assertions for example, a 70% chance that chest pain with elevated troponin indicates myocardial infarction supporting risk prediction, disease progression modeling, and clinical decision support. Tools include Pr-OWL, BayesOWL, and OntoBayes. Fuzzy reasoning handles imprecise concepts like “high temperature” or “moderate risk”; for instance, ICU monitoring may classify “slightly low oxygen saturation” as critical when combined with other signs. Tools include FuzzyDL, Fuzzy OWL 2, and FiRE, with applications in personalized medicine, lifestyle-based risk assessment, and wearable device analytics. Hybrid approaches combine probabilistic and fuzzy reasoning for multi-sensor fusion, managing heterogeneous and uncertain IoT-generated healthcare data [87] [88].

## 3. Research Strategy

The state of the art forms a critical foundation for research of this nature, and a comprehensive literature review must encompass all studies that contribute to ontology-driven healthcare analytics. The scope of this work is to examine the dynamics of Big Data applications, the challenges they present, and the role of semantic technologies in supporting interoperability and decision-making in healthcare systems. Section 3.1 introduces the research questions that emerge from this investigation and outlines how they are addressed throughout the review.

## 3.1 Research Questions

The formulation of research questions is central to shaping this review, as they guide the systematic exploration of ontology-driven Big Data Analytics in healthcare. Given the heterogeneity of healthcare data, the evolving role of semantic technologies, and the importance of knowledge-driven methods, the following research questions frame the scope of this study:

**Q1.** What are the key challenges in applying Big Data Analytics in healthcare?
**Q2.** How do ontologies improve the accuracy and efficiency of data analytics in healthcare?
**Q3.** What are the most commonly used ontologies for healthcare data analytics?
**Q4.** How do ontology-driven approaches enhance predictive analytics and clinical decision-making?
**Q5.** How can knowledge modelling techniques strengthen ontology-driven healthcare analytics?
**Q6.** What role do rule-based systems play in advancing Big Data Analytics for healthcare decision-making?
**Q7.** What are the future trends for ontology-based Big Data analytics in healthcare?

Together, these questions provide the overall structure for the review. The discussion focuses on technical, organizational, and ethical barriers including scalability, interoperability and data quality. It then explores how semantic frameworks contribute by offering standardized vocabularies, enabling interoperability across heterogeneous datasets, and enhancing analytical precision through reasoning. The review also highlights widely adopted ontologies, such as SNOMED CT, ICD, LOINC, and UMLS, emphasizing their role in ensuring semantic consistency and supporting large-scale data integration. Further, it examines how ontological knowledge facilitates early diagnosis, personalized treatment, and decision-support systems. Related approaches such as knowledge graphs, conceptual models, and semantic representations are discussed for their ability to capture complex clinical relationships.

The analysis also considers rule-based inference engines integrated with ontology-driven big data frameworks, which enable automated reasoning, guideline enforcement, and real-time event processing. Finally, attention is given to integration with AI/ML techniques, real-time IoT-driven analytics, adaptive ontology evolution, and interoperable healthcare ecosystems. Importantly, the literature search strategy (Section 3.2) directly aligns with these guiding questions, as the search terms were specifically designed to capture evidence addressing each dimension.

### 3.2 Literature Search Strategy

A structured, multi-step strategy is adopted to ensure comprehensive coverage of literature on ontology-driven Big Data Analytics in healthcare. The process began with the selection of established academic databases widely used in health informatics and computer science, namely IEEE Xplore, PubMed, ScienceDirect, SpringerLink, ACM Digital Library, and Scopus. These platforms collectively capture interdisciplinary research spanning healthcare, semantic technologies, and data science.

Search queries were constructed using keywords in combination with Boolean operators to balance precision and recall. The core terms included "ontology-driven *healthcare analytics", "Big Data in healthcare", "semantic interoperability", "predictive analytics in healthcare", "clinical decision support ontologies", "knowledge modelling in healthcare", "healthcare knowledge graphs" , "rule-based decision support", "complex event processing in healthcare", "semantic data integration", "healthcare data standardization"*, and *"AI and ontologies in healthcare"* Synonyms and related expressions (e.g., *"ontology-based", "semantic frameworks" ,"rule-based reasoning"*) are also incorporated.

Citation chaining is applied to capture seminal contributions and influential recent works not retrieved through keyword searches. To ensure contemporary relevance, the review primarily focused on publications from the last decade, with foundational older works included when offering critical methodological or theoretical insights. Both peer-reviewed journal articles and high-quality conference proceedings are considered.

Table 1 summarizes the structured search strategy, mapping core areas, scenarios, and properties with representative search terms. This design ensures direct alignment between the literature retrieved and the research questions in Section 3.1, such that every query systematically supports the exploration of challenges, ontologies, knowledge modelling, rule-based systems, and future trends.

**Table 7. Searching strategy**

| Area | Scenario | Property | Search Terms |
|---|---|---|---|
| Big Data in Healthcare | Healthcare Data | Challenges | "Big Data challenges" OR "Healthcare scalability" OR "Data privacy" OR "Interoperability" |
| Semantic Technologies | Healthcare Data | Interoperability | "Semantic interoperability" OR "Ontology integration" OR "Data standardization" |
| Biomedical Ontologies | Clinical Applications | Ontology Repositories | "Clinical decision support ontology" OR "SNOMED CT" OR "ICD" OR "LOINC" OR "UMLS" |
| Knowledge Modelling | Representation | Graph-based Approaches | "Knowledge modelling" OR "Healthcare knowledge graphs" OR "Ontology representation" |
| Rule-based Systems | Decision Support | Inference & Reasoning | "Rule-based decision support" OR "Event-condition-action rules" OR "Complex event processing" |
| Future Research Directions | Innovation & Trends | AI and IoT Integration | "AI and ontologies" OR "Ontology predictive modelling" OR "IoT healthcare analytics" |

## 3.3 Selection Criteria of Articles

The following criteria were applied to determine whether the selected articles should be included in this review study.

**Inclusion Criteria**

- Inclusion Criteria 1: Only papers focusing on ontology-based or ontology-driven approaches in healthcare analytics were considered.
- Inclusion Criteria 2: Studies addressing Big Data challenges such as scalability, interoperability, data quality, and privacy were included.
- Inclusion Criteria 3: Articles describing knowledge modelling, healthcare knowledge graphs, or semantic frameworks were included.
- Inclusion Criteria 4: Studies employing rule-based systems, inference engines, or complex event processing for healthcare decision-making were included.
- Inclusion Criteria 5: Peer-reviewed journal articles and conference proceedings published primarily in the last decade were considered, with seminal older works included where relevant.
- Inclusion Criteria 6: Papers presenting empirical validation, case studies, frameworks, or practical implementations relevant to healthcare analytics were included.

**Exclusion Criteria**

- Exclusion Criteria 1: Articles discussing generic Big Data methods without healthcare-specific applications were excluded.
- Exclusion Criteria 2: Studies using ontologies only in a theoretical context, without practical application to healthcare, were not included.
- Exclusion Criteria 3: Works unrelated to knowledge modelling, semantic interoperability, or rule-based decision support were omitted.
- Exclusion Criteria 4: Non-peer-reviewed materials, such as editorials, opinion pieces, or grey literature, were not considered.
- Exclusion Criteria 5: Duplicate studies or incomplete papers lacking sufficient methodological or experimental details were excluded.

By applying the research questions, search strategy, and selection criteria outlined in Section 3, a structured body of literature was identified. Section 4 builds upon this foundation by synthesizing the selected studies into an ontology-integrated framework for healthcare data analytics.

## 4. Ontology-Integrated Framework for Healthcare Data Analytics

The proposed framework integrates ontology with big data analytics to address the heterogeneity, scalability, and interoperability challenges in healthcare. It is designed as a layered architecture that begins with diverse data sources such as EHRs, imaging, IoT devices, and public health records, followed by ingestion, storage, and processing layers to manage both batch and real-time data. Advanced data analytics techniques, including machine learning and semantic reasoning, are then applied to extract actionable insights, which are presented through visualization and decision-support tools for clinical use [78][89]. The ontology layer ensures semantic consistency, interoperability, and enriched querying across disparate data sources, thereby enhancing decision-making and predictive analytics. The layered structure of this ontology-integrated framework is illustrated in Figure 3.

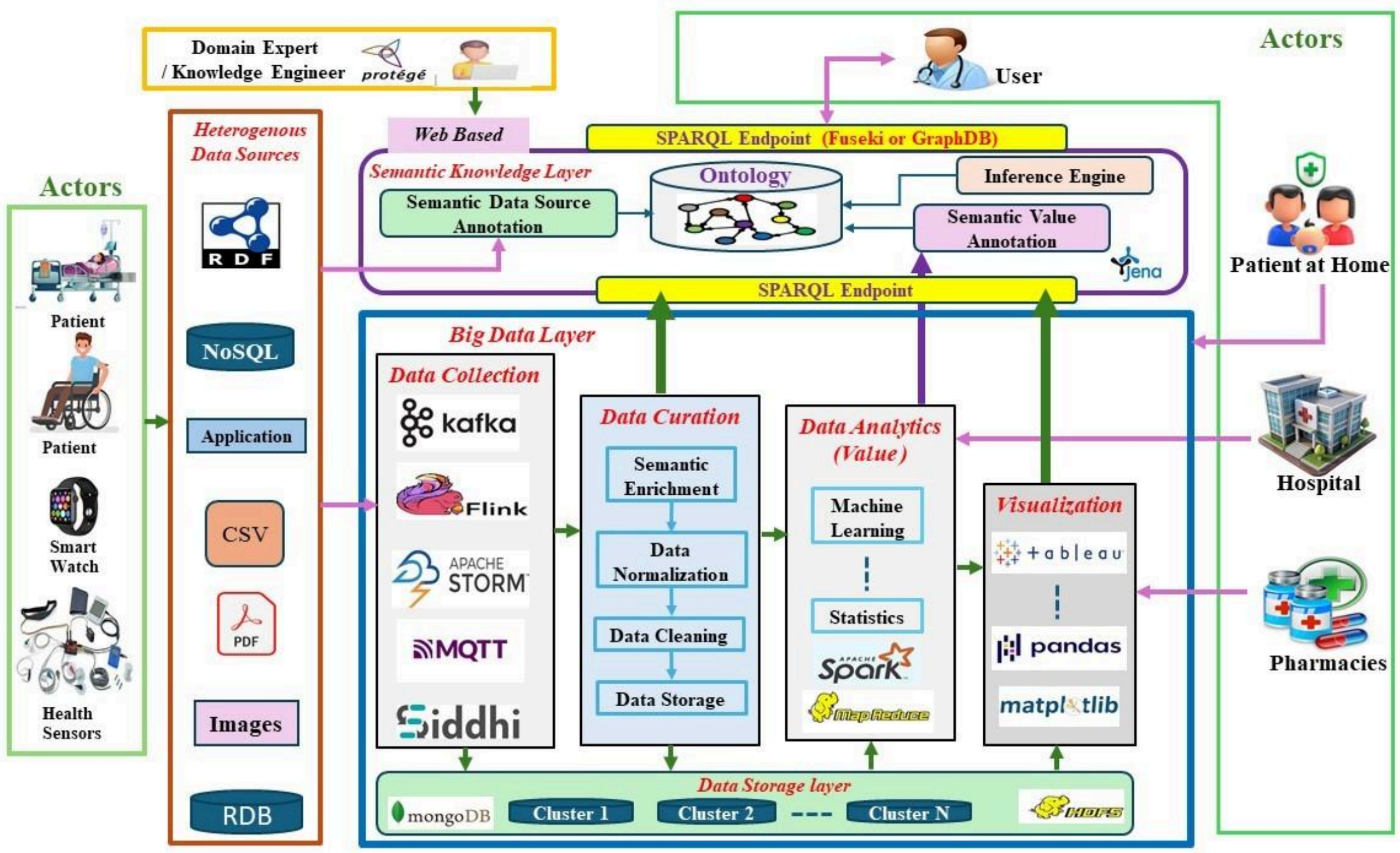

Figure 6. Ontology-driven architecture for healthcare big data analytics

4.1 Data Sources

Healthcare data is derived from heterogeneous sources such as Electronic Health Records (EHRs), clinical images, wearable devices, IoT-based health sensors, and public health repositories. These inputs exist in diverse formats, including relational databases (RDB), NoSQL systems, CSV files, PDFs, and images, making data integration a critical challenge.

4.2 Data Ingestion Layer

The ingestion layer manages the extraction, transformation, and loading (ETL) of data from multiple sources. Streaming tools such as Kafka, Flink, Apache Storm, MQTT, and Siddhi enable both real-time and batch ingestion of patient health data, supporting continuous monitoring as well as historical data processing [90][91][92].

4.3 Data Storage Layer

Once ingested, data is stored in scalable repositories. This includes MongoDB for semi-structured data and clustered storage systems (Cluster 1, Cluster 2 … Cluster N) integrated with Hadoop Distributed File System (HDFS) for large-scale distributed data management. These storage layers act as data lakes and warehouses, accommodating structured, semi-structured, and unstructured healthcare datasets [93][94][95].

4.4 Data Processing Layer

The processing layer facilitates both batch and real-time analytics. Hadoop MapReduce and Apache Spark are employed for large-scale batch processing, while Flink, Apache Storm, and Siddhi support real-time stream processing of IoT and sensor data. This dual approach ensures timely insights while maintaining historical trend analysis [96] [97].

4.5 Analytics Layer

The analytics layer integrates machine learning models, statistical analysis, and semantic reasoning to extract meaningful insights from healthcare data. Processes such as semantic enrichment, data normalization, and cleaning enhance data quality, while predictive analytics aids in disease monitoring, diagnosis, and personalized care [98] [99].

4.6 Visualization and Decision Support

Visualization and decision support tools such as Tableau, Pandas, Matplotlib etc. transform raw analytical outputs into interactive dashboards, clinical reports, and decision-support systems. These tools assist healthcare professionals, patients, and institutions in making data-driven medical decisions and improving treatment outcomes [100][101][102].

4.7 Integration with Ontology Layer

The ontology layer ensures semantic interoperability across diverse healthcare datasets. Using SPARQL endpoints (Fuseki or GraphDB), Jena, and inference engines, the system performs semantic annotation and reasoning shown in Figure 7. This enables unified querying, knowledge discovery, and consistent interpretation of health information across hospitals, pharmacies, and patient homes [103].

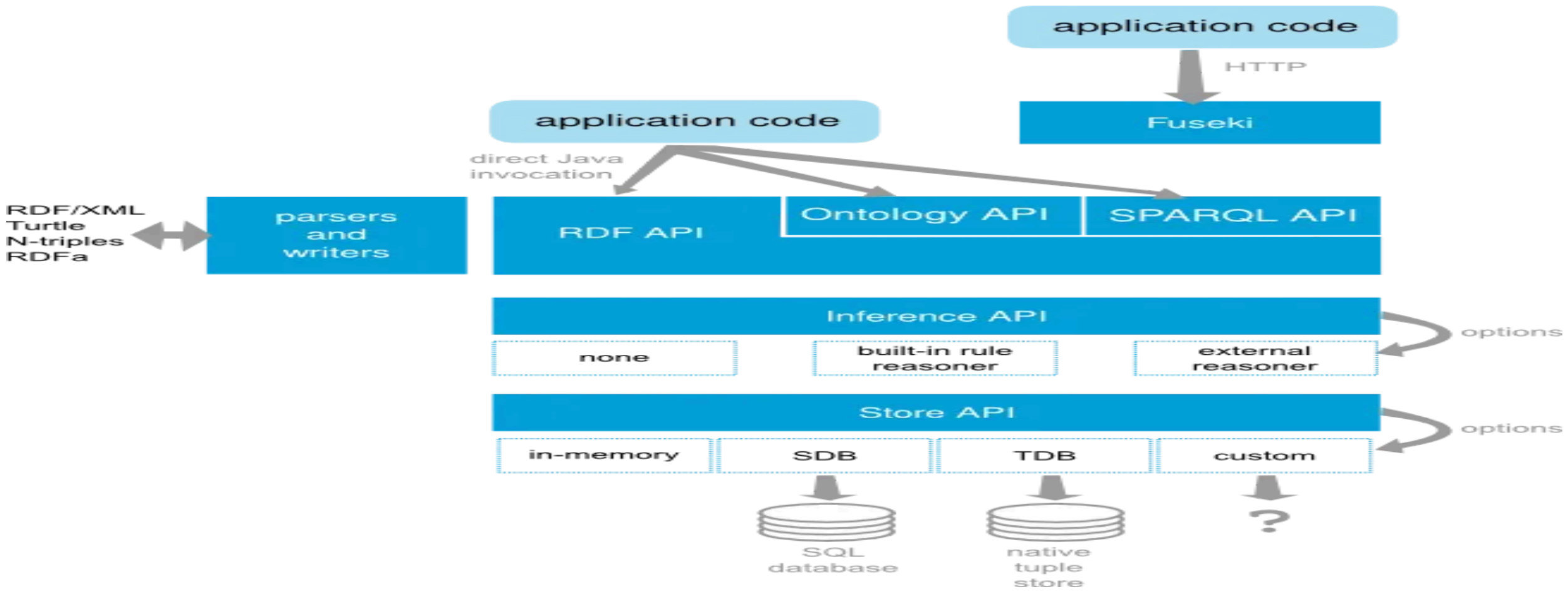

Figure 7. Architecture of the Apache Jena Framework[6]

[6] https://miro.medium.com/1*ToGJc57S_EcPSKRkIjsKmw.png

### 4.8 Ontology to Big data continuous analytical pipeline

In Figure 8, show the overview in practical scenario of Ontology to Big data continuous analytical pipeline is a system designed for real-time data analysis. It starts with the OWL Ontology which is created by a Knowledge Engineer and Domain Expert to define the data structure and relationships. This ontology is then configured by a Data Engineer and mapped by Ontop to a DB-descriptive ontology, which facilitates the translation of SPARQL queries into SQL. This process enables a Data Scientist/Analyst to query the system using the SPARQLWrapper[7] with analytical libraries. At the core of the pipeline is SIDDHI CEP, which leverages FlinkSQL to execute complex queries over both streaming and stored data. Real-time processing is supported by streaming tables and CEP within Flink, while the Data Lake acts as the central repository, continuously populated through data ingestion pipelines involving Spark, Kafka, Flink, and Hive. These components work together to enable real-time processing, pattern detection, and a continuous analytical feedback loop [104].

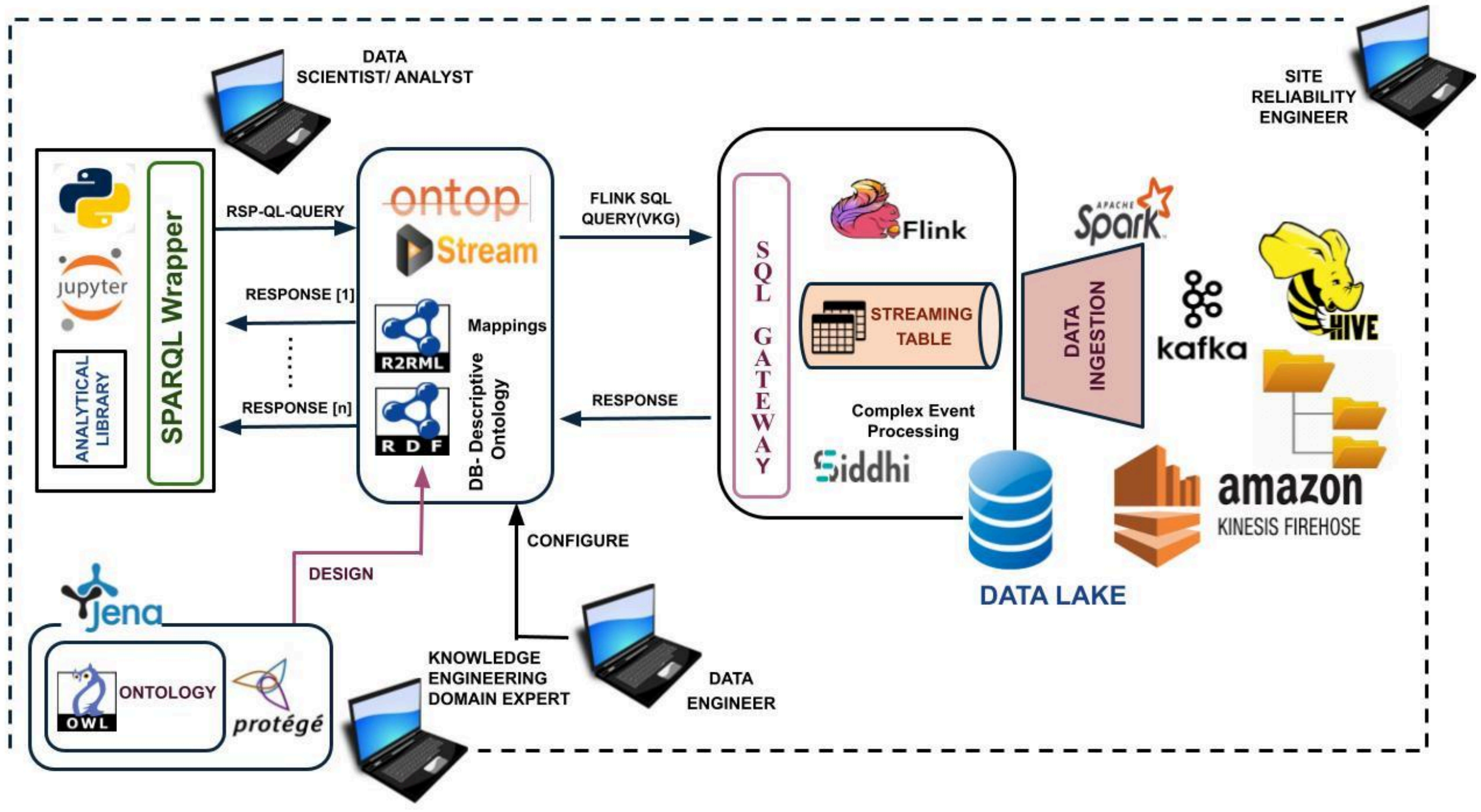

Figure 8. Ontology-Driven Continuous Analytics Pipeline for Big Data[8]

## 5. Ontology-Driven Approaches for Big Data Analytics

This section outlines ontology-driven approaches for big data analytics in healthcare and addresses major challenges such as semantic heterogeneity, interoperability, scalability, and reasoning complexity.

### 5.1 Semantic Techniques for Big Data Analytics

[7] https://github.com/RDFLib/sparqlwrapper
[8] https://chimera-suite.github.io/

Semantic techniques form the methodological basis of ontology-driven analytics. Rather than focusing on data volume alone, these techniques leverage ontologies to embed domain knowledge into healthcare data pipelines. They ensure accurate interpretation of diverse datasets and enable advanced analysis at scale. The core techniques include ontology-based data integration, annotation, validation, and enhanced querying.

### 5.1.1 Ontology-Based Data Integration for Analytics

Ontology-based data integration follows a structured process that begins with data preprocessing [78][105][106]. At this stage, healthcare data such as vital signs, laboratory results, symptoms and conditions, activity and lifestyle records, and biometric information are cleaned and normalized with the help of lookup tables. The processed data is then evaluated for quality, and local schemas are aligned with the domain ontology to achieve semantic consistency [107]. User queries pass through this ontology layer to generate entity lists and combine entities and attributes, while missing values are handled through imputation. Together, these steps create a unified and reliable data source for analytics shown in Figure 9.

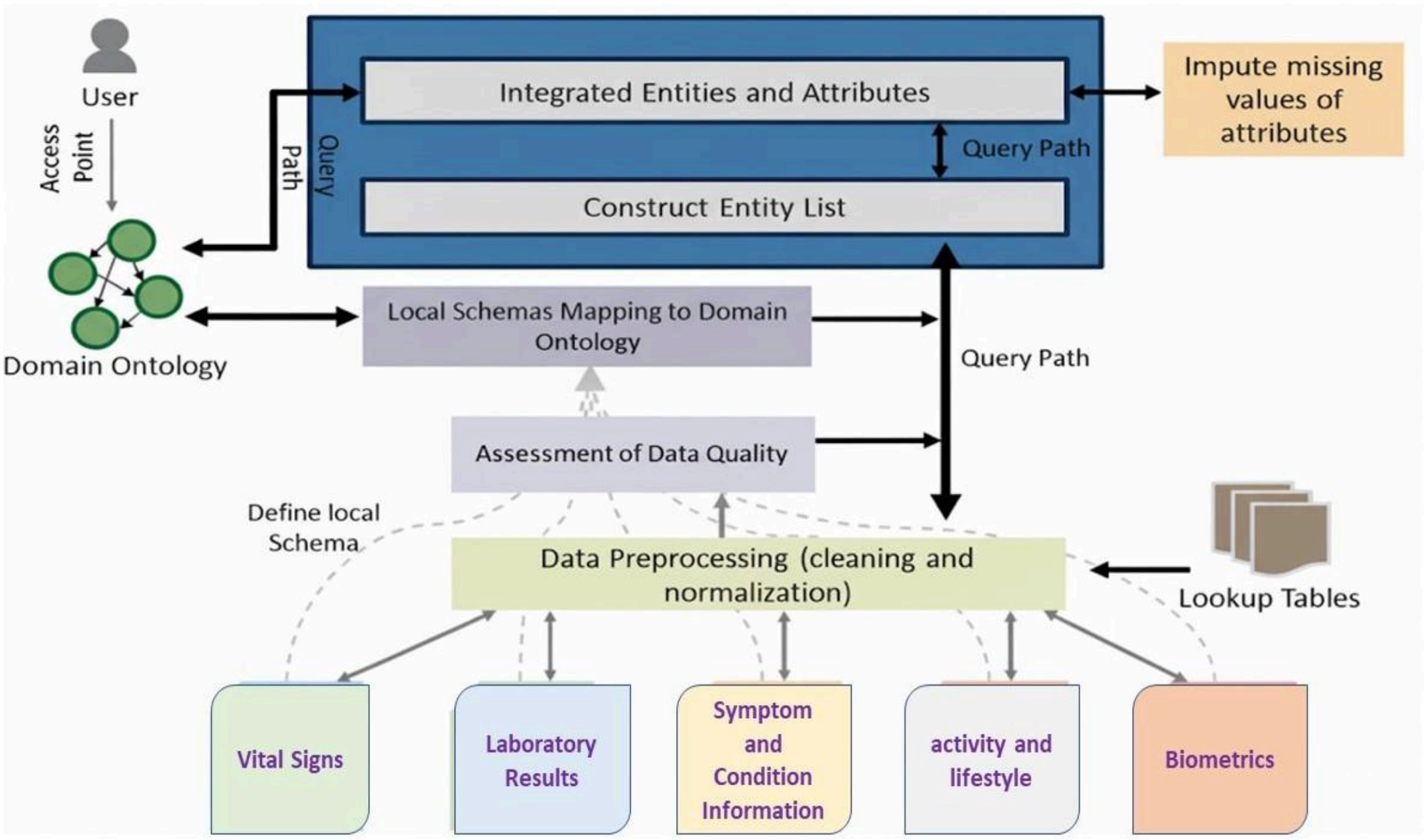

Figure 9. Framework of the proposed ontology-based data integration approach for healthcare

### 5.1.2 Ontology-Based Data Annotation and Tagging for Enhanced Insights

The concepts of ontology-based data annotation are exceptionally relevant and transformative for the healthcare domain, which is characterized by vast, complex, and often unstructured data. Medical information, from a patient's narrative in an EHR to a radiologist's notes on an MRI, is a prime example of data that is difficult to analyze in its raw form. By applying an ontology, healthcare systems can move beyond simple, siloed data and create a unified, semantically rich knowledge base, a foundational step for modern data management. This structure connects apparently disparate data points such as a specific symptom mentioned in a doctor's note, a diagnosis code, and a prescribed medication to a shared, formal understanding of the medical world [108] [78][109][110].

In practice, an ontology for healthcare would define core concepts like "Patient," "Disease," "Symptom," and "Medication," and crucially, the relationships between them, such as a "Patient" "is diagnosed with" a "Disease," which "is associated with" certain "Symptoms," and "is treated by" a "Medication." For example, when a clinician records that a patient is "experiencing severe headaches and photophobia," an ontology-based system can automatically link these terms to the concept of "Migraine" using a standardized terminology like SNOMED CT and advanced Natural Language Processing tools [111]. This process transforms a simple text entry into a structured, relational data point that can be easily queried and analyzed [112] [113][114].

This level of semantic richness unlocks powerful insights for clinical care and research. For a clinician, it can power sophisticated decision support systems that automatically flag potential drug interactions or suggest relevant diagnostic tests [115]. For researchers, it allows for the analysis of massive, integrated datasets to identify new disease biomarkers, discover novel drug targets, or predict patient outcomes with unprecedented accuracy. By providing a consistent framework across different hospital systems and research institutions, medical ontologies also solve a major challenge in healthcare: interoperability [111]. This enables the seamless sharing and aggregation of data for large-scale studies, ultimately accelerating medical discovery and advancing the field of personalized medicine.

### 5.1.3 Ontology-Driven Querying and Data Retrieval for Analytical Purposes

Ontology-driven querying and data retrieval have emerged as powerful approaches for addressing the challenges of heterogeneous and large-scale data analytics. Systems such as **ATHENA** demonstrate how domain ontologies can bridge the gap between natural language queries and relational databases by semantically translating user intent into executable queries, enabling effective analytical querying without requiring technical expertise [116]. Similarly, ontology-based models like **BIGOWL4DQ** extend this paradigm by incorporating data quality reasoning, ensuring that analytical results are based on reliable and semantically validated information [117]. For distributed and domain-specific contexts, the Ontology-Driven Domain Scientific Data Retrieval Model (DSRM) leverages ontologies to represent both queries and scientific datasets, thus facilitating seamless integration and retrieval across heterogeneous sources [118].

Conceptually, ontology-mediated data access has been studied extensively in relation to query expressibility, complexity, and rewritability, particularly through logical frameworks such as Disjunctive Datalog and Constraint Satisfaction Problems [119]. Ontology-driven approaches have also been adapted for semantic analytics in RDF environments, where ontologies provide semantics for aggregation and multidimensional views [120]. More recent works illustrate their applicability in integrating heterogeneous learning analytics data [121] and in enhancing information retrieval precision using ontology-enriched document representations [122]. Collectively, these contributions highlight how ontologies not only standardize data representation but also enable advanced, semantically enriched querying and analytical retrieval across diverse domains.

## 6. Applications of Ontology- driven Big Data Analytics in Healthcare

Ontology-driven big data analytics in healthcare has diverse applications ranging from individual-level clinical care to large-scale population health management. These applications demonstrate how semantic technologies enhance interoperability, reasoning, and decision-making across multiple layers of healthcare. The following subsections highlight major areas where ontology-driven approaches have been successfully applied, starting with clinical applications.

### 6.1 Clinical Applications

Clinical applications include Clinical Decision Support Systems (CDSS), predictive analytics for disease diagnosis, personalized medicine and precision health, public health surveillance and epidemiology, as well as population health management.

#### 6.1.1 Clinical Decision Support Systems

Ontology-driven CDSS have gained significant attention in recent years as they leverage structured domain knowledge, reasoning engines, and big data technologies to enhance the quality and effectiveness of clinical care. These systems are designed to integrate heterogeneous clinical data, codify guidelines, and provide semantically aware decision support in the form of alerts, recommendations, or diagnostic assistance. For instance, a systematic review of ontology-based CDSS rules highlighted that while ontologies are widely applied for representing medical knowledge and terminologies, rule reuse across systems remains limited and rule management practices still lack maturity [123]. Similarly, the PITeS-TIiSS project [124] developed a personalized ontology-based CDSS for complex chronic patients, demonstrating improved semantic interoperability and tailored care pathways.

Ontology-based approaches have also been applied to improve medication appropriateness in older multimorbid patients by modeling drug regimens, sedative load, and potential adverse interactions, ultimately enhancing prescription safety and decision support [125]. More recently, new methods such as active learning pipelines have been proposed to automatically identify and incorporate candidate terms into CDSS ontologies, addressing the challenge of continuously updating knowledge bases with evolving clinical evidence [126]. Collectively, these works illustrate the benefits of ontology-driven CDSS, including semantic interoperability across diverse data sources, improved rule management and maintainability, personalization of patient care, and adaptability to new knowledge. However, limitations persist, particularly in terms of the lack of standardized rule reuse, challenges in real-time large-scale reasoning performance, and barriers to clinician acceptance and integration into existing workflows [127].

#### 6.1.2 Predictive Analytics for Disease Diagnosis

Ontology-driven predictive analytics provides facilities to integrate statistical/ML models, rules, or reasoning engines to predict disease onset, progression, or diagnosis. By embedding domain concepts (symptoms, risk factors, disease taxonomy, lab results, imaging, etc.) into an ontology, these systems can improve interpretability, handle semantic heterogeneity, enrich features, and facilitate integration of multimodal data.

Recent work demonstrates various advantages, such as "*The Impact of Ontology on the Prediction of Cardiovascular Disease Compared to Machine Learning Algorithms"* showed that augmenting ML classifiers with ontology-based features improves accuracy, precision, recall etc., in cardiovascular disease prediction, outperforming many pure ML approaches [128]. Similarly, "*A Decision Support System for Liver Diseases Prediction"* integrates decision-tree-derived rules, SWRL rules, SPARQL queries, and ontology representations to detect liver disease types given clinical data, delivering richer diagnostic suggestions [129]. Another example is "*Ontology-based knowledge representation for bone disease diagnosis"*, which proposes a multimodal deep learning architecture guided by bone disease ontology; it combines imaging, lab, and clinical data to build a diagnosis-support system that maintains interpretability through the ontology structure [130]. Moreover, "*An Efficient Ontology Based Chronic Disease Diagnosis Model"* presents a semantic web-based framework that diagnoses chronic diseases, showing how semantic technologies help in early detection in low-resource settings [131].

Some existing works also use unstructured radiology reports. "*Ontology-driven Text Feature Modeling for Disease Prediction using Unstructured Radiological Notes"* shows how combining clinical ontologies with word embeddings from radiological text can predict disease groups even without structured EHRs, outperforming baseline structured-data models [132]. These works highlight several benefits of ontology-driven predictive disease diagnosis: enhanced interpretability (ontology informs why a disease is predicted), improved feature engineering (ontology helps structure risk factors), ability to integrate heterogeneous data (lab, image, unstructured text), and in some cases better performance than purely statistical models. However, challenges remain: availability of large, high-quality annotated datasets; latency or computational cost when reasoning/inference over large ontologies; generalization across populations; and integrating predictive systems into clinical workflows in a way clinicians trust.

6.1.3 Personalized Medicine and Precision Health

Personalized medicine (or precision health) refers to tailoring healthcare decisions and interventions to the individual patient's biological, environmental, lifestyle, and phenotypic characteristics. Ontology-driven approaches support this by providing formal, machine-readable domain knowledge (ontologies, knowledge graphs, phenotype ontologies etc.) that enable semantic integration of diverse data sources, interpretability, and consistent representation of patient-specific features. For instance, the "Ontology-based modeling, integration, and analysis of heterogeneous clinical, pathological, and molecular kidney data for precision medicine" [133] integrates clinical, pathological, and molecular kidney data with ontologies such as Human Phenotype Ontology (HPO), Cell Ontology (CL), and Uberon, creating a precision medicine metadata ontology (PMMO) to harmonize variables across domains. This facilitates biomarker discovery, phenotype stratification, and supports individualized treatment insights. A common representation schema over varied biomedical data sources (clinical notes, genomics, literature, imaging etc.) and uses semantic integration to generate actionable patient-specific knowledge for diagnosis and treatment (e.g. for cases in dementia and lung cancer) [134]. Additionally, "*Patient-Centric Knowledge Graphs: A Survey of Current Methods, Challenges, and Applications"* highlights how knowledge graphs centered on individual patients combine ontologies, structured and unstructured data, reasoning and inference to provide a holistic patient view supporting precision interventions [135]. In Lifestyle and Wellness, *AI and semantic ontology for personalized activity eCoaching in healthy lifestyle recommendations* uses ontologies together with meta-heuristic optimization to recommend personalized activity plans, illustrating how precision health need not always be about disease but also proactive wellness management [136].

When combined, these applications demonstrate that ontology-driven personalized medicine enhances data interoperability across domains (clinical, molecular, environmental), supports phenotypic and biomarker stratification, improves decision support with greater explainability, and enables proactive, individualized wellness and disease management. Nevertheless, several challenges remain: safeguarding privacy, managing missing or noisy data across sources, ensuring the scalability of ontology reasoning, updating ontologies as biomedical knowledge advances, and integrating such systems into clinical workflows in ways that are both acceptable and safe for clinicians and patients.

### 6.1.4 Public Health Surveillance and Epidemiology

Ontology-driven approaches are increasingly used in public health surveillance and epidemiology to support standardized, timely, and interpretable monitoring of disease outbreaks, risk factors, and population health trends. These methods address issues like heterogeneous data sources (clinical, laboratory, environmental, IoT), inconsistent coding, delayed detection of outbreaks, and poor interoperability among surveillance systems. For example, the development of the COVID-19 application ontology by the RCGP Research and Surveillance Centre (RCGP RSC) enabled reliable case identification, health outcomes tracking, microbiological sampling, and national‑level dashboarding, coping with the changing terminology and coding during the pandemic in primary care settings [137]. Recently, *IoT-MIDO* [138], designed an ontology to bridge individual patient monitoring (including IoT sources), clinical management and infectious disease surveillance enabling risk analysis, early warning, and transforming real‑time patient data into public health cues. The *Drug Abuse Ontology* [139] harnessed web-based data to support epidemiology research related to substance use, illustrating how ontologies can facilitate real-time surveillance of social and behavioural public health challenges. In addition, the *Genomic Epidemiology Ontology* (GenEpiO[9]) offers a controlled vocabulary for infectious disease surveillance and outbreak investigations. It supports a harmonized representation of genomic, epidemiology, and clinical laboratory data, which is critical for fast response to emerging pathogens.

These applications show several important benefits: improved semantic interoperability across surveillance networks; more consistent and interpretable case definitions; earlier detection of disease trends; more fine-grained epidemiological insights (e.g. combining clinical, laboratory, and environmental indicators); and enhanced ability to respond to epidemics in real time or near real time. However, adopting ontology-driven surveillance also brings challenges: ensuring frequent updates to ontologies in face of novel pathogens; managing data privacy and ethical issues especially in IoT and personal health monitoring; computational performance when integrating large, streaming, environmentally-linked datasets; and integrating with public health workflows and policy making where delays, legal/regulatory constraints, or resource limitations may affect adoption.

### 6.1.5 Population Health Management and Analytics

Population Health Management (PHM) and Analytics involve the aggregation, analysis, and management of health outcomes across large groups of individuals, often defined by geography, demographics, or risk factors, to improve overall public health, reduce inequalities, and guide policy and resource allocation. Ontology-driven approaches support PHM by enabling semantic integration of diverse data sources (clinical, environmental, socio-economic, lifestyle), facilitating stratification of populations, improving interpretability of analytic results, and providing frameworks for governance, privacy, and reusability.

[9] https://genepio.org/?utm_source=chatgpt.com

Recent studies illustrate these contributions. For example, the **Social Determinants of Health Ontology (SDoHO)** [140] formalizes key social, environmental, and economic factors and relations among them, showing strong coverage in clinical notes and survey data; this helps in measuring and analyzing how non-medical factors influence health outcomes across populations. The Pathling[10] tool offers analytics on FHIR-formatted data, integrating rich terminologies such as SNOMED CT, to support cohort selection, exploratory data analysis, and predictive modelling over large populations [141]. Another work, *Data Analytics for Health and Connected Care: Ontology, Knowledge Graph and Applications* (DAHCC)[142] and Population health management through human phenotype ontology with policy for ecosystem improvement" [143] defines ontology, knowledge graphs and an ecosystem approach using the Human Phenotype Ontology (HPO) respectively for capturing and integrating sensor metadata, AI model outcomes, patients' health conditions, genomic, phenotypic, environmental, and behavioral data at a national scale in connected care settings. These works collectively show how ontology-based PHM can support stratifying risk groups, tracking population-level health determinants, enabling predictive alerts, and facilitating policy design with explainable evidence.

However, several challenges remain in PHM analytics when using ontology-driven methods: ensuring data privacy and governance across jurisdictions; dealing with data sparsity or bias (especially in social determinants); the computational cost of reasoning over very large knowledge graphs or ontologies; keeping ontologies updated with evolving medical, environmental, and social knowledge; integrating PHM analytics outputs into healthcare systems and policy workflows; and ensuring the explainability of results so that stakeholders (clinicians, public health officials, communities) can trust and act on them.

### 6.2 Ontology and Big Data Toolchains

Ontology and big data toolchains form the technical backbone that enables practical implementation of ontology-driven healthcare analytics. These toolchains consist of diverse software ecosystems that support ontology creation, management, integration with big data platforms, and downstream healthcare analytics. While ontologies provide the semantic layer for knowledge representation and reasoning, toolchains ensure their usability by offering modeling environments, distributed processing platforms, and analytics interfaces. To better understand their roles, this section is divided into three categories: ontology editors that facilitate the design and maintenance of ontologies, big data platforms that enable scalable storage and processing, and healthcare analytics tools that transform raw clinical data into actionable insights.

---

[10] https://pathling.csiro.au/

6.2.1 Ontology Editors

Various tools support ontology development and management in healthcare and big data, spanning open-source, commercial, and research prototypes. Protégé offers scalable ontology modeling (e.g., SNOMED CT) with community support but a moderate learning curve. DeepOnto integrates ontologies with deep learning for disease prediction and drug interaction analysis, though performance may lag on large ontologies. OnSET provides visual semantic search for public health surveillance but lacks advanced reasoning. ODK ensures automated pipelines and quality control for large biomedical ontologies, requiring technical expertise. FluentEditor enables collaborative guideline modeling with a GUI but is limited for very large datasets. DynDiff tracks ontology changes to support evolving healthcare knowledge, complementing other editors. Collectively, these tools facilitate ontology-driven analytics and clinical decision support (see Table 8).

**Table 8:** *Ontology Development and Management Tools for Healthcare and Big Data, with license, scalability, usability, strengths, limitations, and use cases.*

| Tool | License | Scalability | Usability (Expert / Non-Expert) | Key Strengths | Limitations | Healthcare / Big Data Use Case |
|---|---|---|---|---|---|---|
| Protégé[11] | Open-source (GPL) | High (handles large ontologies with plugins) | Expert-friendly; moderate learning curve for non-experts | Mature, community support, rich plugins (reasoners, visualization, SPARQL). | Not very intuitive for beginners; limited built-in collaboration. | Biomedical ontology modeling (SNOMED CT, OBO ontologies), clinical knowledge integration. |
| DeepOnto [12][144] | Open-source (Python package) | Moderate (depends on Python ML stack) | Expert-focused (Python devs, ML researchers) | Bridges ontologies + deep learning; supports embeddings, alignment, taxonomy. | New project; may face performance issues with huge ontologies. | Disease prediction, drug interaction analysis, ontology-based ML pipelines. |

[11] https://protege.stanford.edu/
[12] https://github.com/KRR-Oxford/DeepOnto

| | | | | | | |
|---|---|---|---|---|---|---|
| OnSET [145] | Research prototype (free) | Moderate (scales well for medium-size graphs) | Non-expert friendly; visual UI | Semantic search, interactive ontology exploration, subgraph prototyping | Limited editing features; not designed for large KG reasoning. | Public health surveillance, epidemiology dashboards, exploratory analytics. |
| Ontology Development Kit (ODK) [146] | Open-source | High (automated pipelines, batch operations) | Expert / Maintainer-focused | Templates, validation, quality control, reproducibility, CI/CD integration. | Requires technical setup; not for casual users. | Large biomedical ontology curation (Human Phenotype Ontology, clinical data integration). |
| FluentEditor[13] | Commercial (with academic license) | Moderate (best for medium-sized ontologies) | Highly non-expert friendly (GUI, natural language-like syntax) | Supports OWL2, SWRL, SPARQL; collaborative editing; diagrams + CNL (Controlled Natural Language). | Commercial; not ideal for very large datasets. | Clinical guideline modeling, diabetes care pathways, collaborative CDSS authoring. |
| DynDiff [147] | Open-source (academic) | High (tested on large biomedical ontologies) | Expert-focused | Detects and classifies ontology changes, impact analysis for versioning. | Not an editor; only complements editing tools. | Monitoring evolving healthcare ontologies, updating CDSS rules as evidence changes. |

6.2.2 Big Data Platforms for Analytics

Several big data platforms and tools support healthcare analytics by enabling storage, processing, and semantic enrichment of large-scale medical data, as shown in Table 9. Hadoop provides distributed storage and batch processing for large EHR archives, with ontology-based query rewriting, but suffers from high latency [148]. Apache Spark enables in-memory analytics and ML pipelines with ontology-driven feature engineering, offering fast processing but high memory demands. Kafka + Flink support real-time stream processing and semantic event detection for disease monitoring, though reasoning can add latency [149]. KNIME offers user-friendly workflow-based analytics with semantic preprocessing but has limited scalability [150]. HealthShare enables cross-hospital data integration with ontology-aligned vocabularies, while Innovaccer Datashop[14] provides population health insights and ontology-based terminology normalization, though both are proprietary and costly. Collectively, these platforms facilitate scalable, ontology-driven healthcare analytics.

[13] https://www.cognitum.eu/semantics/FluentEditor/?utm_source=chatgpt.com

Table 9: Big Data Platforms for Healthcare Analytics, showing core functions, healthcare use cases, ontology roles, pros, and cons.

| **Platform / Tool** | **Core Function** | **Healthcare Use Case** | **Ontology Role** | **Pros** | **Cons** |
|---|---|---|---|---|---|
| Hadoop (HDFS, MapReduce, Hive, HBase) | Distributed storage & batch processing | Storing & querying large EHR archives, longitudinal patient datasets | Schema mapping, ontology-based query rewriting in HiveQL | Highly scalable, flexible schema-on-read, widely adopted | High latency, not ideal for real-time analytics, complex setup |
| Apache Spark | In-memory parallel computing | Predictive analytics, ML on patient or genomics data | Ontology-driven feature engineering, semantic enrichment in ML pipelines | Fast iterative processing, rich APIs, MLlib integration | High memory usage, native ontology reasoning limited |
| Apache Kafka + Flink | Real-time ingestion & stream processing | Disease outbreak detection, IoT/sensor monitoring | Ontology-driven Complex Event Processing, semantic event patterns | Low-latency, scalable, fault-tolerant | Ontology reasoning adds latency, complex deployment |
| KNIME | Workflow-based analytics & ML | Clinical decision support, small-to-mid datasets | Ontology nodes for semantic preprocessing, concept mapping | User-friendly GUI, rapid prototyping, good connectors | Limited scalability for very large datasets, GUI-dependent |
| HealthShare (InterSystems) | Health Information Exchange & analytics | Cross-hospital patient data integration | Alignment of vocabularies like SNOMED, LOINC, HL7 | Strong interoperability, real-time aggregation | Proprietary, expensive, limited ontology customization |

[14] https://innovaccer.com/blogs/datashop-8211-the-operating-system-that-powers-healthcare

| | | | | | |
|---|---|---|---|---|---|
| Innovaccer Datashop | Population health & predictive insights | Risk stratification, cohort analytics | Ontology-based terminology normalization | Integrated dashboards, healthcare-focused | Vendor lock-in, costly, limited flexibility |

6.2.3 Healthcare Analytics Tools

Healthcare analytics tools transform large, heterogeneous clinical datasets into actionable insights for patient care, operational efficiency, and population health. They combine data mining, machine learning, visualization, and semantic technologies to support predictive modeling, decision-making, and public health monitoring [1]. Platforms like IBM Watson Health and SAS Health Analytics enable predictive risk modeling and clinical decision support, while open-source tools such as RapidMiner facilitate workflow automation and ontology-driven feature engineering. Visualization tools like Tableau and Power BI provide interactive dashboards for real-time monitoring, and domain-specific platforms like OHDSI/Atlas as shown in Table 10.

Table 10. summarizes key healthcare analytics tools, their core capabilities, typical use cases, and support for ontology or big data integration.

| **Tool/Platform** | **Type** | **Core Capabilities** | **Healthcare Use Case** | **Ontology/Big Data Support** |
|---|---|---|---|---|
| IBM Watson Health [151] | AI-driven analytics | NLP, predictive modeling, clinical decision support | Cancer care, chronic disease management | Uses ontologies for medical concepts |
| SAS Health Analytics [152] | Commercial analytics | Statistical modeling, forecasting, data mining | Risk stratification, outcome prediction | Supports big data integration |
| RapidMiner [153] | Open-source ML | Predictive analytics, data preparation | Disease diagnosis prediction models | Integrates with big data frameworks |
| Tableau [154] | Visualization | Dashboards, trend analysis, real-time monitoring | Hospital performance & patient monitoring | Can integrate with semantic models |
| Power BI [155] | Visualization | Interactive dashboards, real-time reporting | Public health reporting, hospital KPIs | Integrates with ontological schemas |
| OHDSI/Atlas [156] | Domain-specif ic | Standardized analytics on OMOP CDM | Multi-institutional clinical research | Ontology alignment via OMOP/OBO |

### 6.3 Ontology based data access

Ontology-Based Data Access (OBDA) is a transformative paradigm for data integration that provides a high-level, conceptual view of an organization's data through a formal ontology. The core value proposition is to decouple the user from the underlying data, enabling domain experts to query information using a business-friendly vocabulary without needing to know the physical location or structure of the data. This is particularly valuable in modern, heterogeneous environments where information is scattered across a variety of formats, including relational databases, NoSQL databases, and CSV files [157]. At its core, an OBDA system acts as a virtual knowledge graph (VKG) and consists of three foundational components: the Data Layer (the original sources), the Ontology (the conceptual model), and the Mappings (the declarative bridge that links the two) [158]. A critical architectural decision is the choice between materialization, which involves physically moving data, and on-demand query rewriting, which is the cornerstone of the VKG approach. For Big Data, on-demand query rewriting is the only feasible solution because it eliminates the need for data duplication and provides real-time access to information, positioning OBDA as a formal, W3C-backed "semantic layer" that is more agile than traditional data warehousing approaches [159]. This approach has been proven in real-world applications, such as the Statoil (now Equinor) case study, where it allowed geologists to query a massive relational database using their own terminology, reducing the time to get answers from weeks to minutes [160]. The OBDA system automatically translates these high-level SPARQL queries into complex native SQL queries, demonstrating its power in leveraging the strengths of both languages to democratize data access for non-technical users [161] [162].

Despite its compelling benefits, OBDA faces significant challenges. The most critical hurdle is the design-time complexity of creating and managing the mappings, a process that is cumbersome, error-prone, and requires deep subject matter expertise [163]. This complexity represents a major bottleneck for the widespread adoption of OBDA. Another fundamental challenge is the trade-off between the expressive power of the ontology and the need for scalable query performance. OBDA systems rely on the OWL 2 QL profile, a lightweight description logic that restricts expressiveness (e.g., it does not allow for recursion or disjunctive information) to ensure that queries can be efficiently rewritten and delegated to the underlying database engine [164].

Table 11 provides an overview of notable OBDA systems developed between 2015 and 2025, highlighting their data models, query languages, and mapping approaches.

Table 11: Overview of OBDA Techniques (2015–2025)

| Year | System | Data Model | Query Languages | Mapping Model |
|---|---|---|---|---|
| 2015 | Manthey | Relational | SQL / SPARQL | DL-Lite / SQL-to-SPARQL |
| 2016 | Blinkiewicz & Bak | Relational | SQL | Visual OBDA (SQuaRE) |
| 2017 | Mugnier et al. | SQL & Key-Value | SQL, XPath, JSONPath, MongoDB | NO-RL |
| 2017 | OntoMongo | Relational & Document | SQL & MongoDB | Object-relational & Object-Document |
| 2019 | PolyWeb | Relational | SQL | R2RML & RML |
| 2019 | OnTop over MongoDB | Relational & Document | SQL & MongoDB | JSON-to-RDF & SQL-to-RDF |
| 2019 | Ontario | RDF & Relational | SQL | RDF-MT |
| 2019 | Squerral (SANSA) | Relational & NoSQL | Spark- & Presto-SQL | RML+FNO |
| 2019 | Fathy et al. | Labeled Property Graph | Cypher | xR2RML |
| 2020 | Obi-Wan | Relational & Document | SQL & MongoDB | (G)LAV view-based query rewriting |
| 2021 | OnTop4theWeb | REST (CSV, JSON, XML) | SPARQL | R2RML |
| 2021 | Chimera | Relational (Hive) | SparkSQL | R2RML |
| 2021 | OntoCB | Document | Couchbase (N1QL) | Object-oriented |
| 2023 | ForBackBench | Relational & Document | SPARQL & SQL | Mapping translation framework |
| 2023 | LUBM4OBDA | RDF & Relational | SPARQL | Benchmarking mappings |
| 2024 | SEDAR | Data Lake | SPARQL | Semantic Modeling |
| 2024 | ODIN | Relational & NoSQL | SPARQL | Semantic Modeling |
| 2025 | OntoProx | Relational | SQL | DL-Lite / advanced mappings |

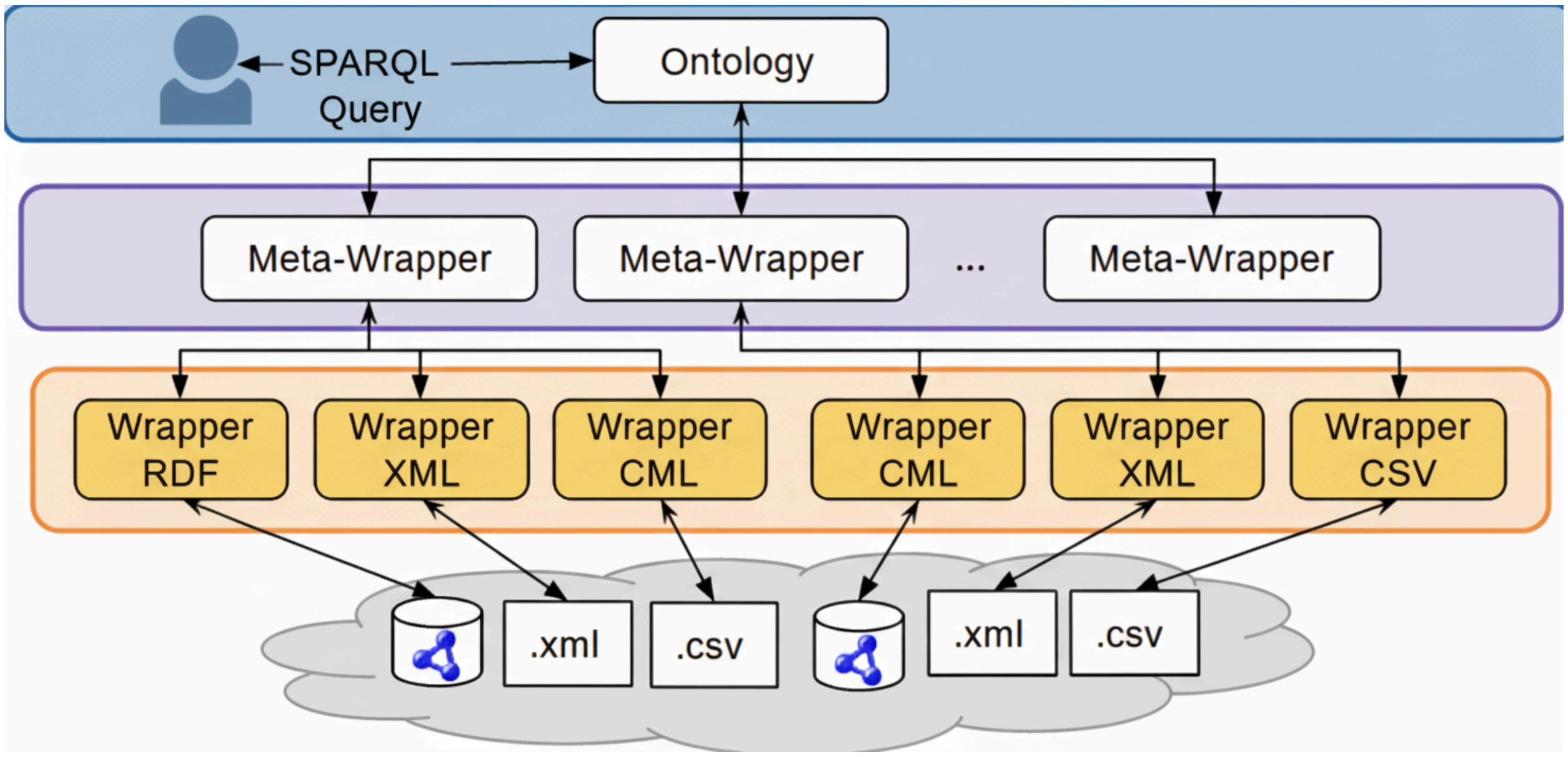

Figure 10. OBDA provides a unified virtual view of heterogeneous data sources through a conceptual ontology and logical mappings.

As illustrated in **Figure 10**, the multi-layered architecture enables users to query heterogeneous data sources through a unified conceptual ontology. When a query is submitted to a SPARQL endpoint, it is handled by different wrappers—such as a Meta-Wrapper for the Data Lake and dedicated wrappers for SQL and CSV sources. These wrappers translate the semantic query into native commands, retrieve the relevant data from the underlying systems, and deliver a unified view to the user without requiring physical data migration.

Research is actively addressing these limitations through automation. Recent advancements in machine learning and Large Language Models (LLMs) are being applied to automate the mapping creation process and reduce manual effort [165]. However, the field still lacks a standardized benchmark for properly evaluating and comparing the performance of different OBDA systems on realistic, real-world queries. The LUBM4OBDA benchmark is a positive step forward, extending prior benchmarks with inference and meta-knowledge capabilities [166], while frameworks such as ForBackBench provide testbeds for assessing mapping translations and performance under varying scenarios [167]. The NPD Benchmark, based on the Statoil use case, is also relevant, but a universal standard is still needed. Looking ahead, the future of data integration will likely adopt a hybrid model, where the semantic layer provided by OBDA coexists with and enhances other data architectures, serving as a critical bridge that makes data more accessible, interpretable, and actionable for all users.

### 6.4 Case Studies

Case Study 1 : Ontology-Driven Drug Recommendation with SIMB-IoT

The Semantic Interoperability Model for Big-data in IoT (SIMB-IoT) [168] was designed to overcome the persistent challenge of semantic heterogeneity in IoT-based healthcare systems. Data generated from wearable devices and mobile health applications are often heterogeneous, fragmented, and difficult to integrate into meaningful clinical insights. SIMB-IoT addresses this issue by introducing an ontology-driven semantic interoperability layer that transforms raw device data into standardized, machine-understandable knowledge. Through RDF-based semantic annotation, diverse health symptoms are represented in a unified format, ensuring that data from multiple sources can be consistently stored, retrieved, and analyzed. Ontological mappings explicitly connect symptoms, diseases, drugs, and side effects, thereby resolving challenges of inconsistent terminology, lack of interoperability, and hidden clinical relationships. These mappings allow physicians to explore complex associations, such as identifying drugs effective for multiple conditions while simultaneously predicting potential adverse effects. By enabling SPARQL-based queries over annotated knowledge graphs, the framework delivers validated and explainable drug recommendations enriched with side-effect awareness [169][170]. In this way, SIMB-IoT leverages ontology not only to harmonize heterogeneous IoT health data but also to improve the accuracy, transparency, and efficiency of personalized healthcare decision-making, as shown in Figure 11.

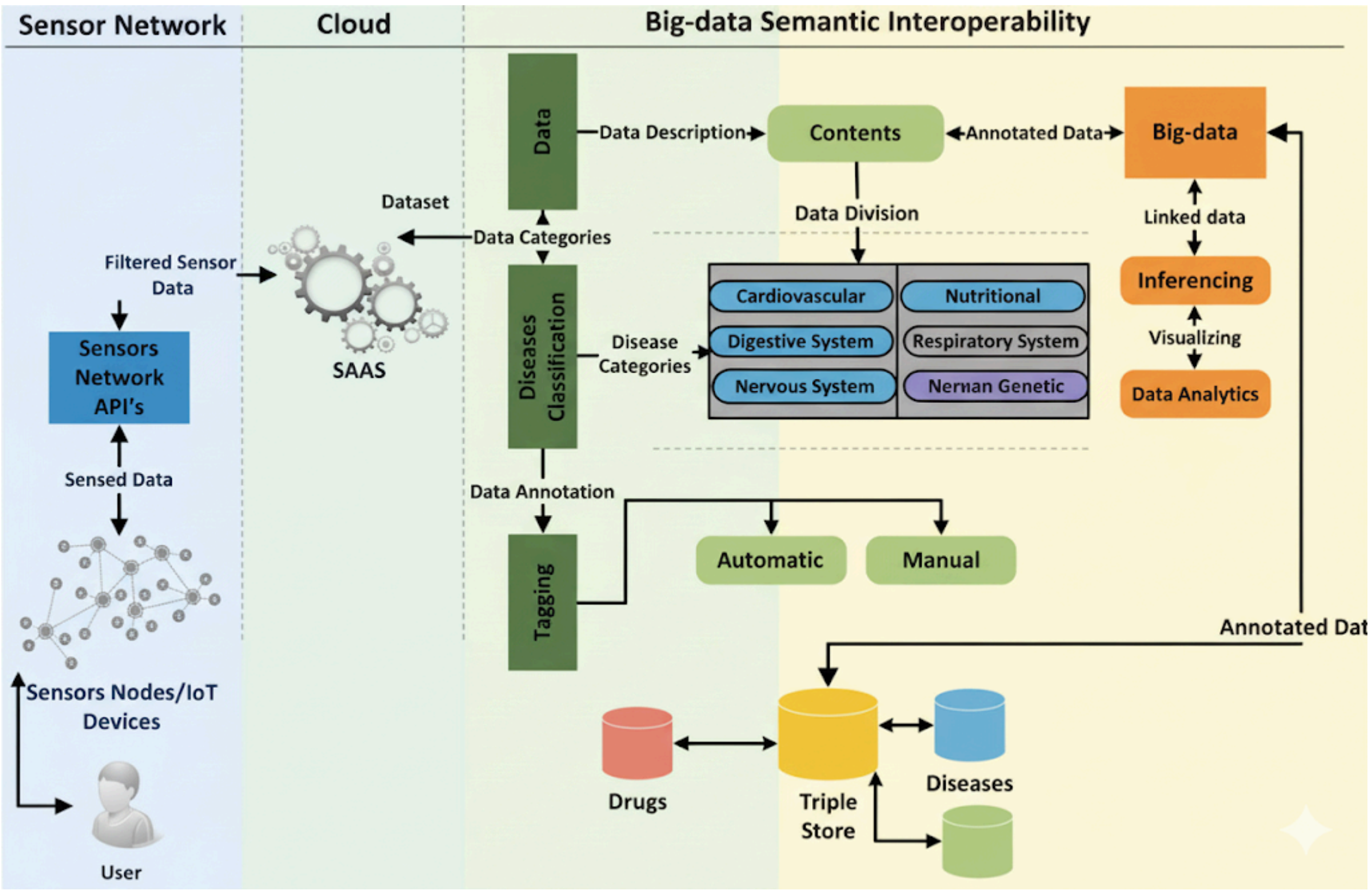

Figure 11: Ontology-Driven SIMB-IoT Framework for Drug Recommendation with Side-Effect Awareness [168]

## Case Study 2. Ontology-Based Schema Design for Big Health Data in NoSQL Databases

In big data scenarios for healthcare, ontologies offer a structured approach to managing the massive volume, variety, and velocity of data commonly recognized as the defining characteristics of Big Data. Traditional relational databases are often inefficient in handling such complex data structures, particularly with respect to query performance. To address this, the paper proposes an ontology-based approach for designing NoSQL databases tailored to semi-structured and unstructured health data. The ontology, serving as a repository of shared and machine-processable knowledge, is leveraged to develop a data model that captures domain semantics, resulting in a schema design that enhances query performance. The effectiveness of this method is demonstrated by comparing the query response times of an ontology-driven NoSQL database against a traditional relational model using the same healthcare data. This approach helps fill the gap in standardized design methodologies for NoSQL databases and emphasizes efficient data retrieval, which is essential for healthcare applications, as illustrated in Figure 12 [73].

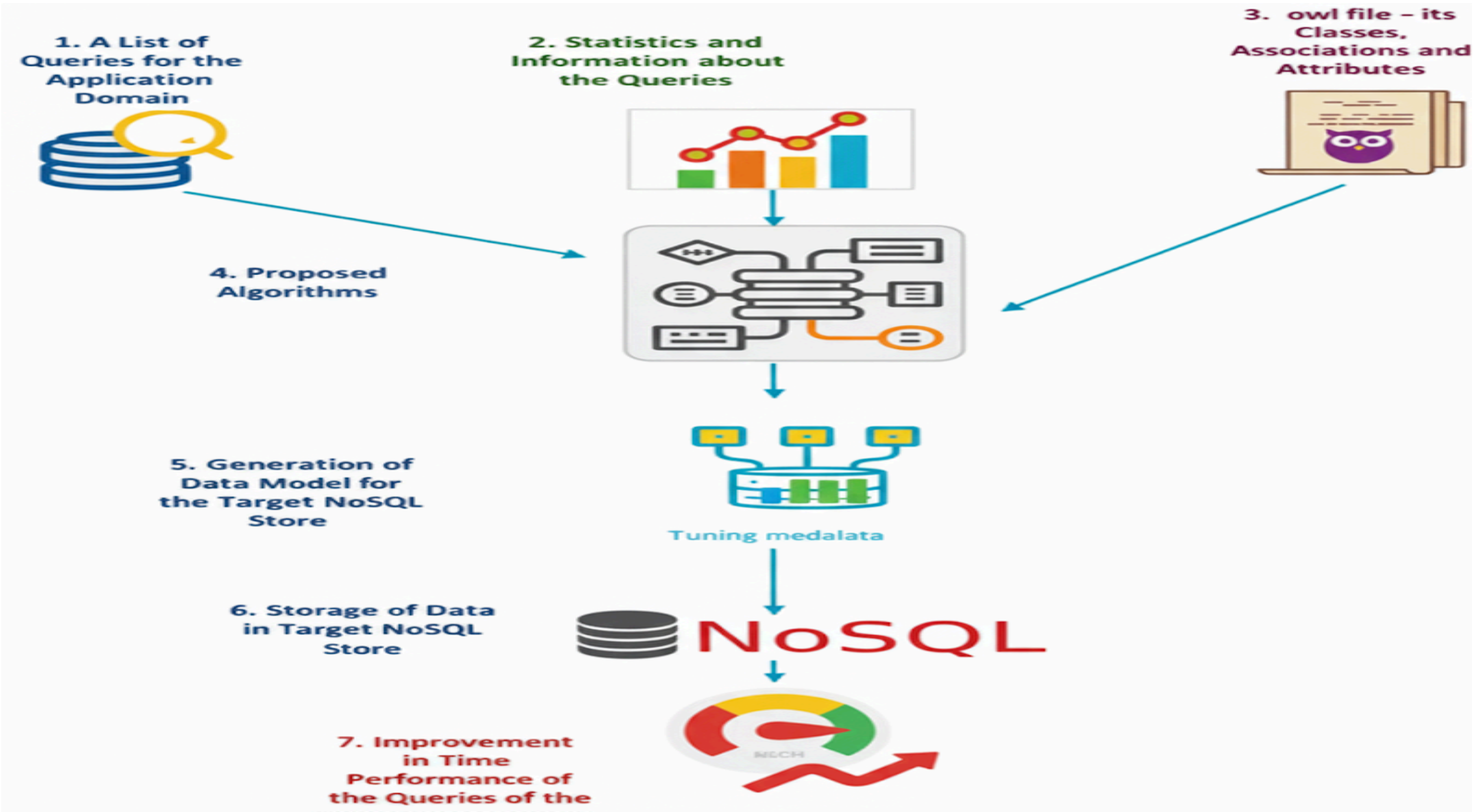

Figure 12. Ontology-Driven Schema Design Workflow for Big Health Data in NoSQL (MongoDB) [73]

## Case studies 3. Ontology-Based Knowledge Modelling in Big data

One use case of ontology-based knowledge modeling in Big Data is found in the healthcare sector, particularly in the integration of medical and oral health data [170]. The complexity and diversity of this data, along with information silos across different health domains, often hinder collaborative patient care and decision-making. To overcome these challenges, semantic web technologies such as ontologies are employed to process, integrate, and share information at a semantic level, ensuring that the meaning of the data is preserved. This approach supports decision-making capabilities such as alerts, recommendations, and explanations enabling healthcare professionals to analyze shared and interdependent knowledge from both domains and deliver more comprehensive, informed patient care [171][172].

## Case studies 4. Ontology based complex event processing in Big Data

Ontology-based Complex Event Processing (OCEP) addresses the challenges that traditional CEP systems face in Big Data environments, namely semantic heterogeneity and data interoperability. The proposed framework, which uses ontologies for reasoning and the RDF to organize event data, improves knowledge-driven event reasoning and decision-making. Implemented within a Hadoop environment, the system utilizes HDFS for scalable storage and Apache Kafka for real-time event execution. A real-time healthcare case study, which uses IoT sensor data to monitor illnesses, serves as a use case that validates the OCEP framework's ability to improve early disease detection and aid in decision-making by integrating multiple event streams with 85% accuracy [173] shown in Figure 8.

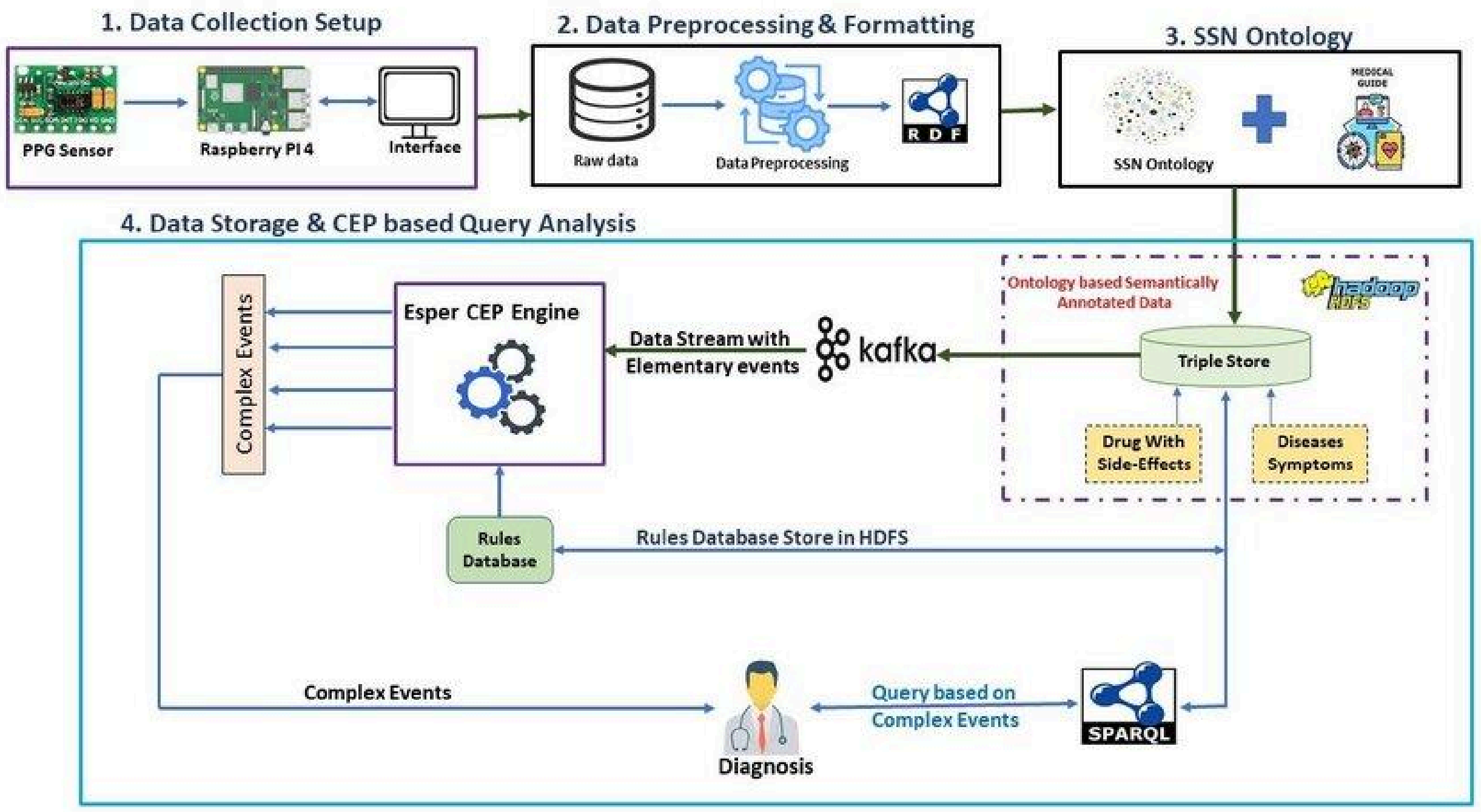

Figure 8. A Framework for Real-time Health Monitoring and Diagnosis using Complex Event Processing [173]

## 7. **Challenges and Evaluation**

This section highlights the key challenges, which include overhead and usability issues, technical interoperability and abstractions, applicability within big data frameworks, benchmarking and evaluation, as well as automation through AI and LLMs.

### 7.1 Key Challenges

Ontology-driven Big Data analytics holds enormous promise for healthcare, but its deployment in real-world systems is far from straightforward. Several technical, organizational, and ethical hurdles must be overcome to achieve reliable, scalable, and sustainable solutions. These challenges span performance constraints, lack of standardized data integration, privacy and security risks, and the need for continuous ontology updates. Ontology-driven Big Data analytics in healthcare faces several interconnected challenges, detailed in the following subsections.

**Initial Overhead and Usability:** Generating meaningful semantic models and knowledge graphs for heterogeneous datasets requires significant time, resources, and domain expertise. Even with pre-existing

KGs, creating accurate mappings to underlying data sources is labor-intensive. Current automated approaches reduce effort but still require substantial human input.

**Evaluation and Benchmarking:** The accuracy of automatically generated semantic labels and models is often validated only on limited benchmark datasets. Initiatives like SemTab and VC-SLAM provide benchmarks for semantic labeling and model standardization, yet more comprehensive evaluation frameworks are needed to assess real-world performance.

**Technical Interoperability:** Most semantic modeling approaches are optimized for relational or tabular data, leaving NoSQL and other heterogeneous sources under-supported. Modern OBDA research emphasizes support for multiple query languages, federated query processing, and compliance with W3C Semantic Web standards to ensure interoperability.

**Human-in-the-Loop and Technical Abstraction:** Even with advanced AI techniques, human oversight remains essential for verifying and refining semantic models and mappings. Enhanced user interfaces that abstract technical complexity and guide non-technical users are critical to maintaining model quality and usability.

**Applicability in Big Data Environments:** Tools like Squerall and Chimera show promise for OBDA in Big Data scenarios but face limitations in query expressiveness, mapping complexity, and platform dependencies. Broader development and community support are needed to generalize these solutions for heterogeneous Big Data systems.

**Leveraging AI/LLMs:** Large Language Models (LLMs) such as ChatGPT offer opportunities to automate ontology creation, semantic mapping, and data integration. Early studies indicate AI can handle complex tasks in knowledge graph generation and data integration, though human oversight remains necessary to ensure reliability and accuracy.

### 7.2 Evaluation Metrics and Benchmarks

**Table 12. Dimensions, Metrics, Tools, and Challenges in Ontology-Driven Healthcare Analytics**

| Dimension | Metric | Purpose | Typical Tools / Benchmarks | Challenge Addressed |
|---|---|---|---|---|
| Accuracy | Coverage | Capture all key healthcare concepts and relations | SNOMED CT, LOINC, ICD-10 | Incomplete domain knowledge |
| | Consistency | Ensure logical soundness and avoid contradictions | Pellet, HermiT, FaCT++ | Ontology conflicts, risk of misdiagnosis |
| Scalability | Query Latency | Measure average query execution time | SPARQL endpoints (GraphDB, Virtuoso) | Real-time performance bottlenecks |
| | Throughput | Assess system capacity under heavy load | LUBM, BSBM stress tests | High-volume data streams |
| Interoperability | Standards Compliance | Verify adherence to RDF/OWL, HL7-FHIR standards | W3C, HL7 specifications | Lack of data standardization |
| | Semantic Alignment | Integrate heterogeneous datasets | Ontology alignment frameworks | Data silos, poor cross-system integration |
| Data Quality | Completeness & Veracity | Minimize missing or inaccurate data | Gold-standard datasets | Low-quality or inconsistent data |
| | Annotation Precision | Improve ontology-based tagging of unstructured data | NLP + ontology annotators | Complexity of free-text clinical notes |
| Security & Ethics | Access Control | Enforce sensitive-data access policies | Ontology rules, HIPAA/GDPR frameworks | Unauthorized access and compliance risks |
| | De-identification | Preserve privacy while retaining analytical utility | Privacy-preserving tools (anonymization) | Patient confidentiality concerns |
| Clinical Utility | Diagnostic Gain | Enhance prediction and decision-making accuracy | F1, AUC, Precision–Recall metrics | Low diagnostic accuracy in practice |
| | Workflow Fit | Assess clinician usability and adoption | Surveys, pilot studies | Poor system integration/adoption |
| Benchmarks | Datasets & Tools | Ensure reproducibility and fair comparison | MIMIC-III/IV, PhysioNet, Protégé | Lack of standardized benchmarks |

**7.3 Literature Review Statistics**

In this section, we present visualizations of the number of papers cited from different domain applications based on ontology-based approaches for big data analytics. Additionally, we illustrate the distribution of cited articles across various digital libraries. A PRISMA diagram of the articles included in the paper is shown in Figure 9. The distribution of articles by year (2015–2025) is shown in Figure 10, the distribution of articles by digital library in Figure 11, and the distribution of articles by domain in Figure 12.

.

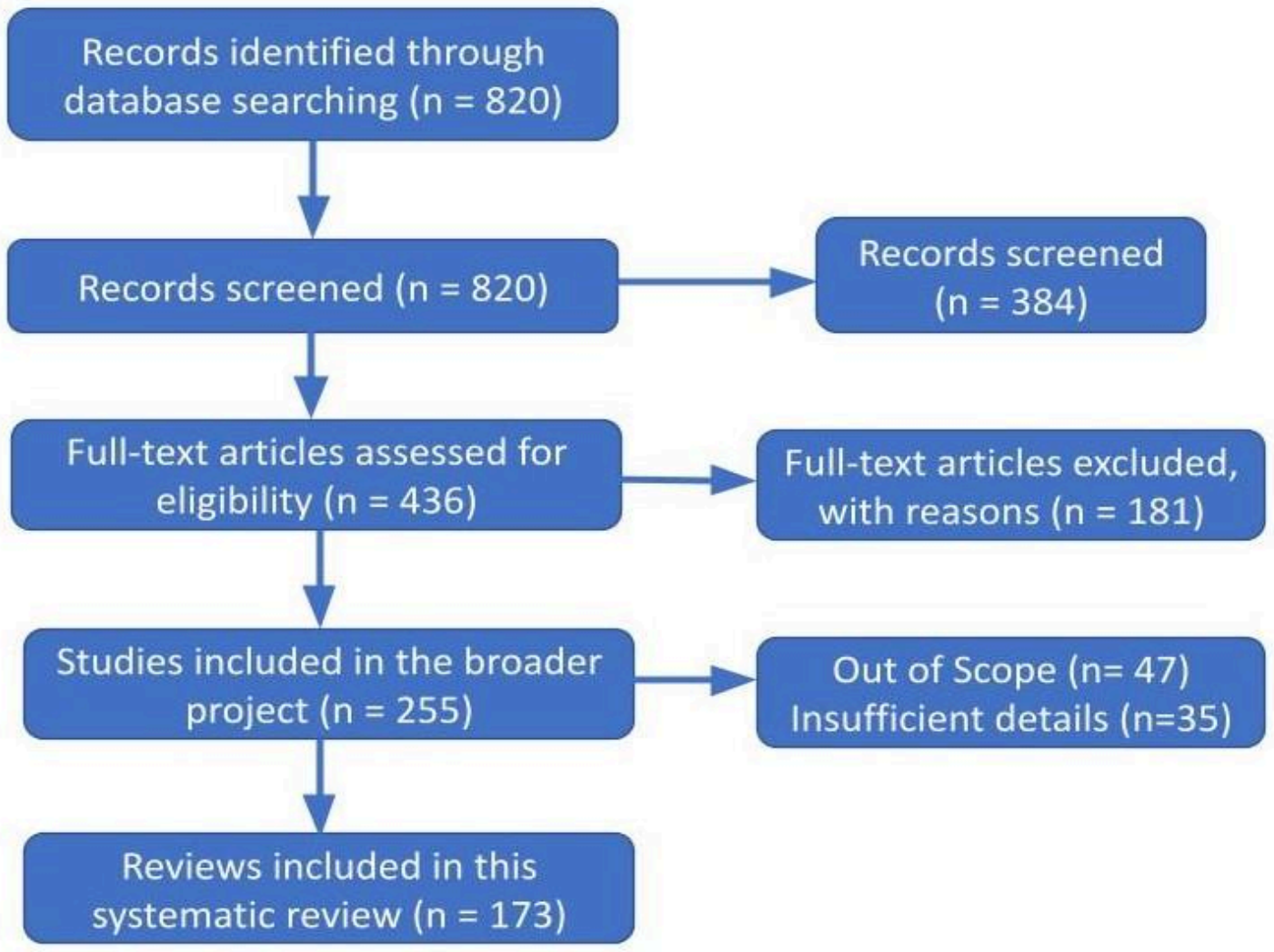

Figure 9. Study flow

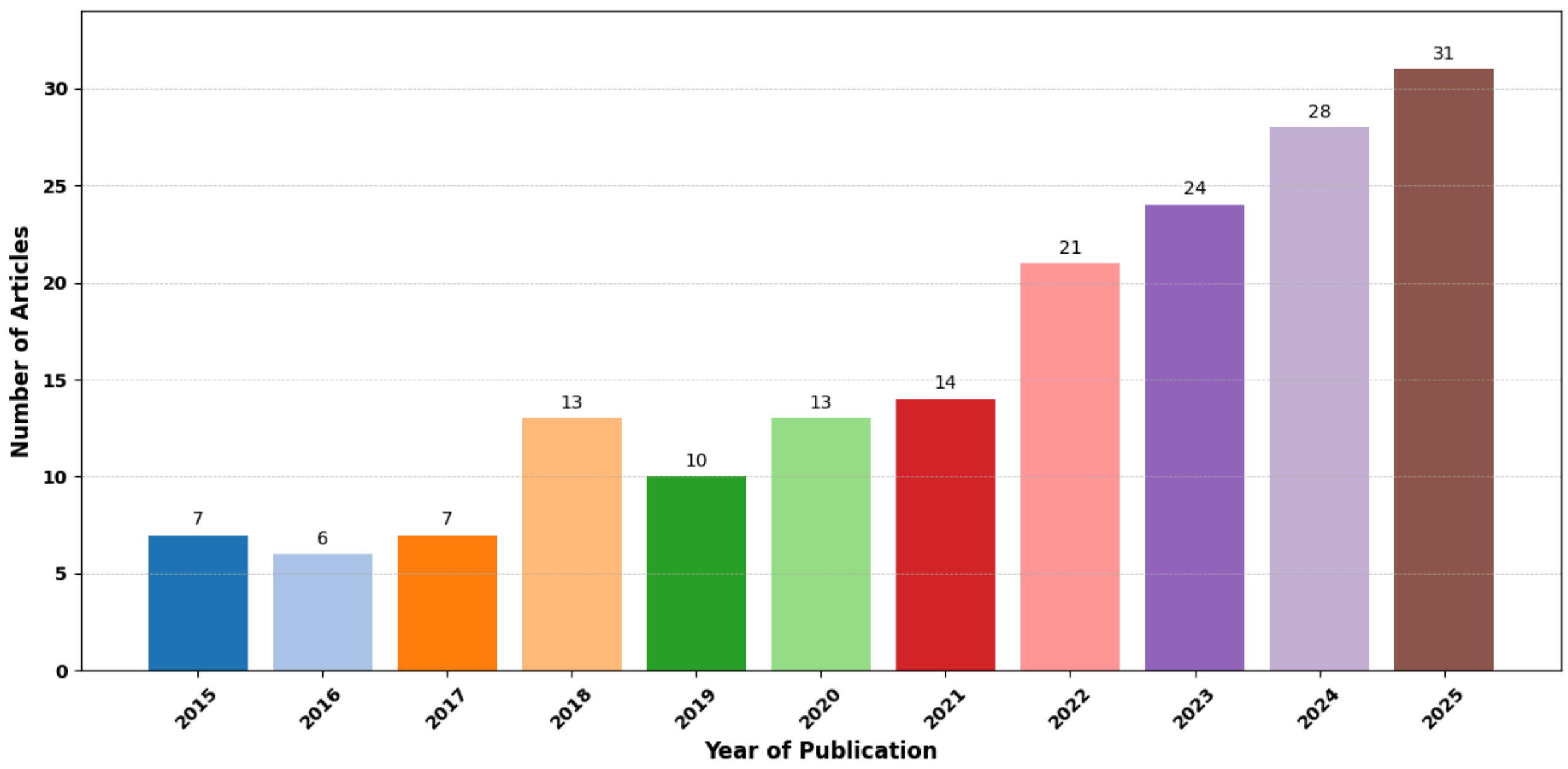

Figure 10. Distribution of articles by year

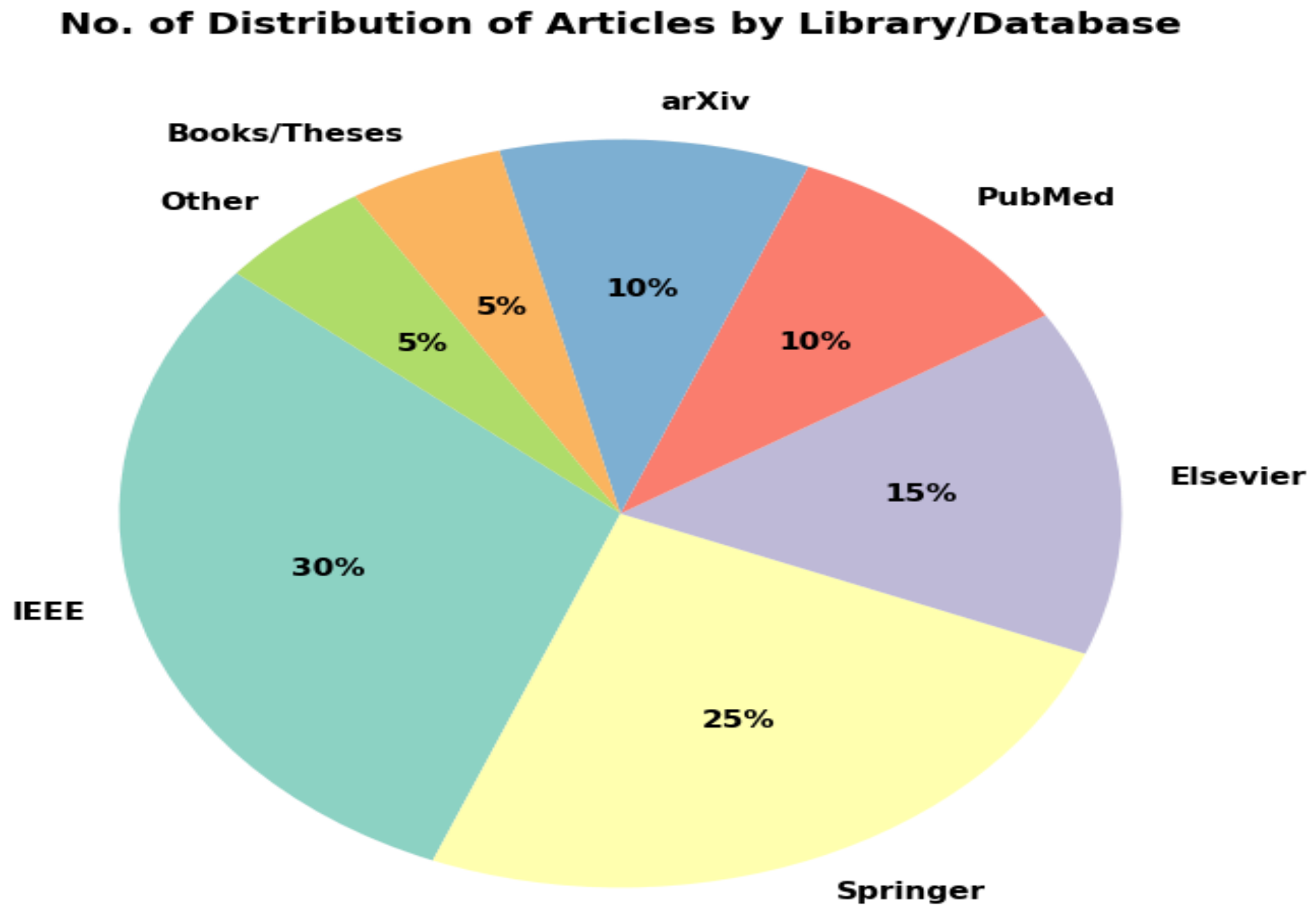

Figure 11. Distribution of articles by library

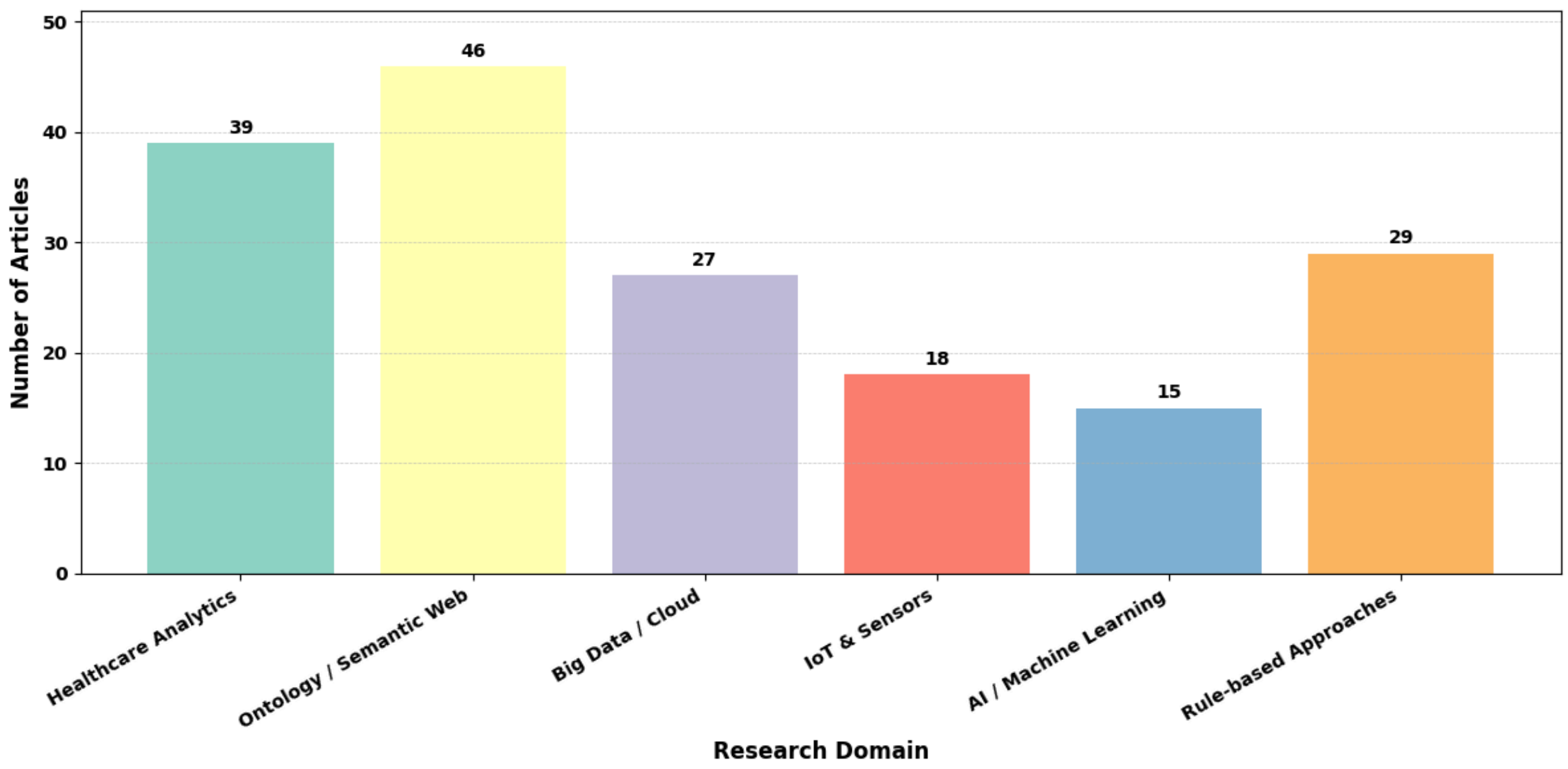

Figure 12. Distribution of articles by domain

## 8. Conclusion and Future Research Directions

This segment includes summarized takeaways from this study and their implications in the healthcare domain, along with future research directions.

### 8.1 Summary of Key Findings

Ontology-driven approaches provide a semantic layer over heterogeneous healthcare data, enabling efficient integration, reasoning, and analytics. OBDA systems leverage ontologies, mappings, and virtual knowledge graphs to simplify access for non-technical users while supporting complex Big Data environments, including Hadoop, Spark, and NoSQL databases. These approaches improve interpretability, interoperability, and accuracy, particularly in clinical decision support, predictive disease diagnosis, personalized medicine, and population health management. Challenges remain in usability, scalability, interoperability, and human-in-the-loop model refinement, highlighting areas for future research.

### 8.2 Implications for Healthcare Analytics

The integration of ontologies with Big Data analytics enhances healthcare decision-making by ensuring consistent interpretation across diverse datasets, enabling real-time insights, and improving predictive and prescriptive analytics. Semantic models facilitate automated data integration, query rewriting, and knowledge reasoning, making healthcare data more actionable for clinicians, administrators, and researchers. These methods also support compliance with data standards, interoperability across systems, and more personalized, evidence-driven healthcare delivery.

**8.3 Future Directions**

Based on the challenges identified in Section 7, the future of ontology-driven Big Data analytics in healthcare should focus on addressing usability, interoperability, and scalability while leveraging emerging AI and IoT technologies:

1. Reducing Initial Overhead and Enhancing Usability: Future work should focus on automating the generation of knowledge graphs (KGs) and semantic models for heterogeneous datasets. Although automated solutions exist, human input remains critical for ensuring accuracy and domain relevance. Enhanced user interfaces that abstract technical complexity will be essential for supporting non-technical users and accelerating adoption.

2. Improved Evaluation and Benchmarking: Standardized benchmarks for semantic labeling, model generation, and query rewriting need to be expanded. Initiatives similar to SemTab[15] and VC-SLAM[16] should be further developed to provide comprehensive testbeds for validating ontology-driven methods under real-world conditions, ensuring reproducibility and accuracy across diverse datasets.

3. Enhanced Technical Interoperability: Research should prioritize support for heterogeneous data sources beyond relational and tabular formats, including NoSQL databases, file systems, and streaming data. Future OBDA systems must enable federated query processing across multiple platforms while maintaining compliance with W3C Semantic Web standards to maximize interoperability.

4. Scalable Big Data Integration: Tools such as Chimera[17] have demonstrated promise for Big Data scenarios, but limitations remain in query expressiveness, mapping flexibility, and platform compatibility. Future research should generalize these solutions to support a wide range of Big Data and NoSQL systems, enabling more scalable and practical deployments.

5. Leveraging AI and LLMs: Recent advancements in AI, including LLMs like ChatGPT, offer opportunities to automate ontology creation, semantic mapping, and data integration. Future research should explore how general-purpose LLMs can be customized for domain-specific data integration tasks, optimizing efficiency while maintaining accuracy. Integrating LLMs into existing workflows could significantly reduce manual effort, improve mapping quality, and handle complex reasoning across heterogeneous healthcare datasets.

---

[15] https://sem-tab-challenge.github.io/2024/
[16] https://live.european-language-grid.eu/catalogue/lcr/7698
[17] https://chimera-suite.github.io/

6. Human-in-the-Loop and Technical Abstraction: Even with advanced AI and automation, human oversight will remain crucial for verifying and refining semantic models and mappings. Developing intuitive, interactive tools that guide users through technical processes will ensure high-quality knowledge representations without overburdening domain experts.

To conclude, this article has provided an overview of semantics-based methods for healthcare data management, integration, and analytics, linking these findings to emerging semantic data lake approaches. Despite notable progress, challenges remain in scalability, usability, evaluation, and human-in-the-loop refinement, reflecting the gap between Big Data platforms, OBDA, and semantic modeling of heterogeneous datasets. At the same time, the convergence of Big Data ecosystems and Semantic Web technologies offers a transformative pathway. By combining distributed frameworks with ontologies, AI, real-time IoT streams, and advanced reasoning, more scalable and adaptive healthcare analytics systems can be developed. Such approaches hold the potential to enhance clinical decision support, enable personalized medicine, and strengthen public health surveillance, ultimately driving more accurate, timely, and patient-centered healthcare decision-making in the years ahead.

## Acknowledgment

The author gratefully for the support received under the National Fellowship for Persons with Disabilities (NFPWD) scheme from the University Grants Commission (UGC), India, for pursuing Ph.D. research. We also thank the Indian Institute of Information Technology, Allahabad, for providing the necessary infrastructure and resources. Special thanks to the Big Data Analytics (BDA) Lab members for their valuable input and support in this entire work.

## Refrences